\documentclass[prb,amsmath,amssymb,twocolumn]{revtex4}
\usepackage{graphicx}
\usepackage{bbm,color,ulem}

\newcommand{\br}{{\bf r}}
\newcommand{\bx}{{\bf x}}
\newcommand{\by}{{\bf y}}

\newcommand{\bp}{{\bf p}}

\newcommand{\bz}{{\bf z}}

\newcommand{\bK}{{\bf K}}

\newcommand{\bR}{{\bf R}}

\newcommand{\ms}{{m^{\ast}}}

\DeclareMathAlphabet{\mathpzc}{OT1}{pzc}{m}{it} 
\begin{document}
\title{Quantum multicriticality in bilayer graphene with a tunable energy gap}
\author{Robert E. Throckmorton}
\author{S.\ Das Sarma}
\affiliation{Condensed Matter Theory Center and Joint Quantum Institute, Department of Physics, University of Maryland, College Park, Maryland 20742-4111 USA}
\date{\today}
\begin{abstract}
We develop a theory for quantum phases and quantum multicriticality in bilayer graphene in the presence
of an explicit energy gap in the non-interacting spectrum by extending previous renormalization group
(RG) analyses of electron-electron interactions in gapless bilayer graphene at finite temperature to
include the effect of an electric field applied perpendicular to the sample, which produces an energy
gap in the single-particle electron-hole dispersion.  We determine the possible outcomes of the resulting
RG equations, represented by ``fixed rays'' along which ratios of the coupling constants remain constant
and map out the leading instabilities of the system for an interaction of the form of a Coulomb interaction
that is screened by two parallel conducting plates placed equidistant from the electron.  We find that some
of the fixed rays on the ``target plane'' found in the zero-field case are no longer valid fixed rays, but
that all four of the isolated rays are still valid.  We also find five additional fixed rays that are not
present in the zero-field case.  We then construct maps of the leading instability (or instabilities) of
the system for the screened Coulomb-like interaction as a function of the overall interaction strength and
interaction range for four values of the applied electric field.  We find that the pattern of leading
instabilities is the same as that found in the zero-field case, namely, that the system is unstable to a
layer antiferromagnetic state for short-ranged interactions, to a nematic state for long-ranged interactions,
and to both for intermediate-ranged interactions.  However, if the interaction becomes too long-ranged or
too weak, then the system will exhibit no instabilities.  The ranges at which the nematic instability first
appears, the antiferromagnetic instability disappears, and the nematic instability disappears all decrease
with increasing applied electric field.  Our main qualitative finding, that the applied electric field
opposes the emergence of symmetry-broken phases in general, suppressing, however, the antiferromagnetc
phase more strongly compared with the nematic phase, is directly testable experimentally.
\end{abstract}
\maketitle
\section{Introduction}
The problem of electron-electron interactions in bilayer graphene has been of great interest
both experimentally and theoretically, especially the problem of what interaction-induced
spontaneous symmetry-breaking orders appear.  In particular, we refer to two recent works\cite{CvetkovicPRB2012, ThrockmortonPRB2012}
by one of the authors where the role of electron-electron interaction in producing novel
emergent quantum phases in gapless bilayer graphene was discussed in great detail using
the renormalization group (RG) theoretical approach.  The current work is a generalization of this previous work\cite{CvetkovicPRB2012, ThrockmortonPRB2012}
to an experimentally relevant situation where the layer symmetry is broken explicitly by an
external electric field so that the non-interacting energy spectrum has an energy gap, and
thus the starting bilayer graphene system is a semiconductor with a gap rather than a
gapless semimetal as in Refs.\ \onlinecite{CvetkovicPRB2012, ThrockmortonPRB2012}.  The original
motivation for studying the possibility of spontaneous symmetry breaking theoretically was twofold.
First of all, near two special points in the Brillouin zone, labeled $\pm\bK$, the electronic
bands make contact with each other at four Dirac-like cones\cite{DasSarmaRMP2011} (trigonal
warping).  If we ignore all but the nearest-neighbor hopping, however, then these cones merge
into parabolic degeneracies.  In two dimensions, this parabolic dispersion results in a logarithmic
divergence of the (non-interacting) susceptibilities to various symmetry-breaking orders at
zero temperature.  Second of all, the fact that the susceptibilities to many different orders
are all divergent makes the problem of determining which will appear once interactions are taken
into account interesting and technically challenging.  When one accounts for trigonal warping,
however, the logarithmic divergences of the susceptibilities will be cut off.  Nevertheless,
symmetry-breaking order is still possible, even with trigonal warping\cite{CvetkovicPRB2012}.

Much theoretical work has already been done on this problem using a variety of methods.  In
addition to numerous mean-field studies\cite{NilssonPRB2006,MinPRB2008,CastroPRL2008,NandkishorePRL2010,NandkishorePRB2010,NandkishoreArXiv,JungPRB2011,MacDonaldPhysScr2012,GorbarPRB2012,ZhangPRL2012,KharitonovPRB2012,KharitonovPRL2012,ZhuPRB2013},
many of which consider the effects of applied electric and magnetic fields, as well as numerical work\cite{LangPRL2012,SchererPRB2012},
a few studies have employed RG techniques as we do \cite{CvetkovicPRB2012,VafekYangPRB2010,VafekPRB2010,LemonikPRB2010,MacDonaldPRB2010,LemonikPRB2012,ThrockmortonPRB2012,FanZhangPRB2012,MurrayPRB2014}.
While most of these consider the problem at zero temperature, two of them\cite{CvetkovicPRB2012,ThrockmortonPRB2012}
treat the problem at finite temperature.  The first of these\cite{CvetkovicPRB2012} is dedicated to
mapping out the possible leading instabilities toward spontaneous symmetry breaking, while the second\cite{ThrockmortonPRB2012}
considers the leading instability for finite-range interactions as a function of overall interaction
strength and range.  This second paper finds that the system is unstable to a layer antiferromagnetic
phase for short-range interactions, in which the electronic spins enter a ferrimagnetic arrangement
in each layer, with the overall polarizations of each layer oppositely directed\cite{LangPRL2012},
to a nematic phase for long-range interactions, in which the rotational symmetry of the system is
broken and the electronic spectrum is reconstructed to have two Dirac cones, and instabilities to both
for intermediate-range interactions.  The short-range result agrees with that obtained in Ref.\ \onlinecite{VafekPRB2010},
and the long-range result with that obtained in Refs.\ \onlinecite{VafekYangPRB2010}, \onlinecite{LemonikPRB2010},
and \onlinecite{LemonikPRB2012}.  In addition, a more recent RG study considers the effect of a finite
chemical potential\cite{MurrayPRB2014}, while all of the other RG studies only considered the half-filled
case, as we will be doing here.

There are also a number of experiments\cite{YacobyScience2010,YacobyPRL2010,Mayorov2011,VelascoNatNano2012,FreitagPRL2012,VeliguraPRB2012,BaoPNAS2012}
that claim to see signatures of symmetry breaking in bilayer graphene.  Most of these observe a gap, but one\cite{Mayorov2011}
performed on an especially clean sample (reported mobility on the order of $10^6\,\text{cm}^2/(\text{V}\cdot\text{s})$)
sees evidence for a nematic phase.  The rest of the experiments, which observe gaps, are inconclusive about the exact
nature of the gapped phase.  Two of these experiments\cite{YacobyScience2010,VelascoNatNano2012} have also looked at
electric field effects and find that the gap closes for a sufficiently strong electric field applied in either direction.

A great deal of both experimental and theoretical work has also been done on various possible quantum Hall phases in
bilayer graphene in the presence of an applied perpendicular magnetic field.  We do not consider or discuss the bilayer
graphene quantum Hall phases at all in the current work since the quantum Hall physics is qualitatively different from
the zero magnetic field situation of our interest because of the quenching of the non-interacting kinetic energy.  All
quantum Hall phenomena are well beyond the scope of this work.

As is implied in this summary of the previous work, the effect of an electric field on symmetry-breaking
orders has been studied in some earlier works experimentally and, at the mean-field level, theoretically.
While the problem has been addressed to a limited extent using RG methods\cite{CvetkovicPRB2012}, we wish
to perform a comprehensive RG-based study of the effect of an electric field on the system.  We therefore dedicate
this work to extending the previous finite-temperature RG work\cite{CvetkovicPRB2012,ThrockmortonPRB2012}
to include the effect of an electric field applied perpendicular to the sample.  In particular, we will
determine what effect the presence of such a field has on the possible leading instabilities that the
system can in principle exhibit, as well as how this field affects the instabilities that the system shows
for a microscopic finite-range electron-electron interaction (i.e., for initial coupling constants corresponding
to a finite-range interaction term present in the original tight-binding model).

Our main reason for embarking on a comprehensive RG study analyzing the effect of an external electric field
on the interaction-driven instabilities in bilayer graphene is the possibility that such an analysis will enable
a direct and unambiguous experimental determination of the quantum phase diagram of interacting bilayer graphene
by carefully studying the experimental data as a function of the applied electric field which is fairly
straightforward to carry out in the laboratory.  Another reason motivating our work is the question of how an
existing energy gap in the non-interacting dispersion affects the quantum criticality in bilayer graphene, in
particular, which symmetry-broken phases are relatively enhanced (suppressed) by this new tuning parameter (i.e.,
the electric field).  Given that earlier works have looked systematically at the effects of interaction range
and strength, finite temperature, finite chemical potential, etc., on the bilayer interaction-driven quantum
instabilities, it makes sense for us to do the same in the presence of an external electric field, which is an
adjustable parameter that can be easily controlled experimentally.

We begin with a tight-binding model of bilayer graphene with a layer energy difference added to represent
the effect of the applied electric field and a density-density interaction term for the electrons, from which
we derive the corresponding effective low-energy theory.  At first sight, it may appear that the
problem becomes intractably difficult because of the explicit breaking of the layer symmetry with many new
fermionic interaction terms getting generated compared with the zero electric field situation.  While, in
principle, the reduced symmetry of our problem would allow for additional four-fermion interaction terms
beyond the nine that are possible without an electric field present\cite{CvetkovicPRB2012,ThrockmortonPRB2012},
it turns out that these additional coupling constants are not generated in our RG flows unless they are
present to begin with, which they are not for any realistic electron-electron interaction.  Therefore, the
problem effectively only has the original nine coupling constants.  We then make use of the same set of RG
equations already derived in Ref.\ \onlinecite{CvetkovicPRB2012} in the presence of an applied electric
field to determine the possible fixed rays that the RG flows can tend to, along with their associated leading
instabilities, as well as to determine the leading instabilities of the system for finite-range (microscopic)
electron-electron interactions through numerical integration of the RG equations.  The fact that the number
of distinct coupling constants in the current problem, although large (9), remains the same as in the previous
work considerably simplifies the theoretical analysis.  We only consider the case of a Coulomb-like
interaction screened by the presence of two parallel infinite conducting plates placed equidistant from the
electron, as this corresponds to the only experimental setup that would allow one to realize a finite electric
field while maintaining the system at half filling, which is precisely the problem that we are interested in: the
quantum phase diagram of interacting intrinsic undoped bilayer graphene in the presence of an external electric
field.

We find that a number of the fixed rays present on the ``target plane,'' a two-parameter family of such fixed
rays found in Ref.\ \onlinecite{CvetkovicPRB2012}, are no longer valid fixed rays, but the four isolated
rays found therein are still valid.  In particular, we find that the region corresponding to the layer-polarized
state, in which there is an imbalance of charge on one layer with respect to the other, is completely excluded,
which is not surprising, considering that that phase no longer breaks any symmetries of the system since the
starting non-interacting system already has a broken layer symmetry.  In addition, we find five other fixed
rays that are not present in the zero-field case, thus making the current problem technically more difficult
than the corresponding zero-field case.  We then construct a map of the leading instability (or instabilities)
of the system as a function of overall interaction strength and interaction range for the screened Coulomb-like
interaction described above for four values of the electric field.  We find that the sequence of instabilities
that one observes is the same as in the zero-field case\cite{CvetkovicPRB2012,ThrockmortonPRB2012}; we see an
antiferromagnetic instability for short-range interactions, a nematic instability for long-range interactions,
and both instabilities for intermediate-range interactions.  However, we find that, if the interaction is too
weak or too long-ranged, then the system will exhibit no instabilities at all, in contrast to the zero-field case.
The ranges at which the nematic instability first appears, at which the antiferromagnetic instability disappears,
and at which the nematic instability disappears all decrease with increasing electric field.  The interaction
strength below which all instabilities disappear, regardless of interaction range, also increases with increasing
layer energy difference.  Thus, a generic effect of the applied electric field is to suppress various interaction-driven
instabilities, which is perhaps not surprising since we expect the electric-field-induced energy gap in the
non-interacting dispersion to act as a cut off on the minimal interaction strength necessary to effect many-body
instabilities.

The rest of the paper is organized as follows.  In Sec. II, we describe the starting model, both the tight-binding
model and the corresponding effective low-energy theory.  Section III is dedicated to explaining our RG analysis and
how we determine the fixed rays of our system; we also state the results for various fixed rays there.  We consider the
case of a microscopic finite-range interaction in Sec. IV, and then present our conclusions in Sec. V.  We give a
number of technical equations and formulas, namely, definitions of various functions and coefficients appearing in the RG
equations, in the Appendices.

\section{Our model}
\subsection{Tight-binding lattice Hamiltonian}
\begin{figure}[tb]
\centering
\includegraphics[width=\columnwidth,clip]{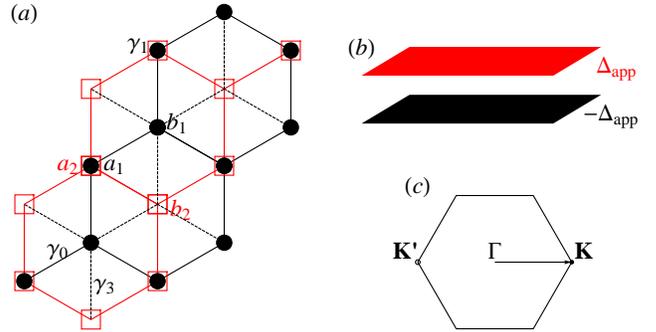}
\caption{\label{Fig:BLGDiagram}(a) The honeycomb bilayer lattice formed by
bilayer graphene.  The black circles represent layer 1 (bottom) and the red
squares layer 2 (top).  We label the dimerized sites $a_i$ and the non-dimerized
sites $b_i$.  The hopping between nearest-neighbor sites in the same layer
is $\gamma_0$, the hopping between the dimerized sites is $\gamma_1$, and the
hopping between nearest-neighbor non-dimerized sites in opposite layers is $\gamma_3$.
(b) Illustration of the two layers of the system, with the energies due to an
applied electric field $\pm\Delta_{\text{app}}$ shown.  (c) The Brillouin zone
associated with the honeycomb bilayer lattice.  We mark the parabolic degeneracy
points $\bK=\frac{4\pi}{3\sqrt{3}a}\hat{\bx}$ and $\bK'=-\bK$.}
\end{figure}
Our starting point is the well-known non-interacting bilayer graphene tight-binding model\cite{DasSarmaRMP2011,NilssonPRB2006,MinPRB2008},
but with an energy difference (or equivalently energy gap) $2\Delta_{\text{app}}$ between
the two layers representing the effect of an applied electric field perpendicular to the
sample.  An illustration of the lattice and the associated Brillouin zone is shown in Fig.\
\ref{Fig:BLGDiagram}.  The Hamiltonian is
\begin{equation}
H=H_0+H_I.
\end{equation}
Here, $H_0$ is the non-interacting part of the Hamiltonian,
\begin{eqnarray}
H_0=&-&\gamma_0\sum_{\bR,\delta,\sigma}(a^\dag_{1\sigma}(\bR)b_{1\sigma}(\bR+\delta)+a^\dag_{2\sigma}(\bR)b_{2\sigma}(\bR-\delta) \cr
&+&\text{H.c.}) \cr
&-&\gamma_1\sum_{\bR,\sigma}(a^\dag_{1\sigma}(\bR)a_{2\sigma}(\bR)+\text{H.c.}) \cr
&-&\gamma_3\sum_{\bR,\delta',\sigma}(b^\dag_{1\sigma}(\bR+\delta)b_{2\sigma}(\bR+\delta+\delta')+\text{H.c.}) \cr
&+&\Delta_{\text{app}}\sum_{\bR,\sigma}\sum_{k=1}^{2}(-1)^k(a^\dag_{k\sigma}(\bR)a_{k\sigma}(\bR) \cr
&+&b^\dag_{k\sigma}(\bR-(-1)^k\delta)b_{k\sigma}(\bR-(-1)^k\delta)), \label{Eq:tbHamiltonian}
\end{eqnarray}
where the $a_{m\sigma}(\br)$ and $b_{m\sigma}(\br)$ operators act on electrons on the
dimerized and non-dimerized sites, respectively, on layer $m$, site $\br$, and with spin
$\sigma$.  The vectors $\bR$ are the positions of the dimerized sites within a unit cell
projected into the $xy$ plane.  The $\delta$ vectors are each one of the three
vectors connecting an $a_1$ site to a nearest-neighbor $b_1$ site, and may take on
the values of $-\frac{\sqrt{3}}{2}a\hat{\bx}+\frac{1}{2}a\hat{\by}$,
$\frac{\sqrt{3}}{2}a\hat{\bx}+\frac{1}{2}a\hat{\by}$, and $-a\hat{\by}$, where $a\approx 1.4\text{ \AA}$
is the lattice constant.  Whenever there is a sum on $\delta$, we sum over these
three values, while we choose one of the three values if $\delta$ appears without a
summation over it.

The second part of the Hamiltonian, $H_I$, is the interaction, given by
\begin{eqnarray}
H_I&=&\tfrac{1}{2}\sum_{k=1}^{2}\sum_{\br\br'}V_{\|}(\br-\br')[n_k(\br)-1][n_k(\br')-1] \cr
&+&\sum_{\br\br'}V_{\bot}(\br-\br')[n_1(\br)-1][n_2(\br')-1].
\end{eqnarray}
Here, $\br$ runs over all lattice sites, once again projected into the $xy$ plane,
and $n_k(\br)=\sum_{\sigma}c_{k\sigma}^\dag(\br)c_{k\sigma}(\br)$, where
$c=a$ or $b$, as appropriate.  We assume that the interaction $V(\br)$ depends
only on distance, i.e., $V(\br)=V(|\br|)$.  In the above expression, we define
$V_{\|}(\br)=V(\br)$ to be the intralayer interaction and $V_{\bot}(\br)=V(\br\pm c\hat{\bz})$,
where $c\approx 3.7\text{ \AA}$ is the distance between the two layers, to be the
interlayer interaction.

Like the system without a layer energy difference, this Hamiltonian
describes an undoped system at half-filling when the chemical potential is zero.
While it is not invariant under a particle-hole transformation alone\cite{ThrockmortonPRB2012},
it is invariant under said transformation followed by an interchange of
the layers.  Using this fact, one may prove that the total occupation number
of the $a$ sites per unit cell is $2$, as is that of the $b$ sites.  Therefore,
while there will be an imbalance of occupation numbers between layers, the
overall system is still at half-filling.

\subsection{Low-energy effective theory}
If we now write the partition function associated with the above Hamiltonian
and integrate out the high-energy modes, or equivalently the dimerized sites, as the
hopping between them sets the splitting of the high-energy bands in the absence of an
electric field\cite{ThrockmortonPRB2012}, we obtain the equivalent low-energy effective
theory, which is
\begin{equation}
Z=\int\mathcal{D}[\psi^\ast,\psi]\,\exp\left (-\int_{0}^{\beta}d\tau\,L_{\text{eff}}\right ), \label{Eq:PartFunc}
\end{equation}
where the Lagrangian $L_{\text{eff}}$ is
\begin{eqnarray}
L_{\text{eff}}&=&\int d^2\br\,\psi^{\dag}\left [\frac{\partial}{\partial\tau}+H(\bp)\right ]\psi\cr
&+&\tfrac{1}{2}\sum_{S\in\mathcal{G}}g_S\int d^2\br\,(\psi^{\dag}S\psi)^2-\mu'\int d^2\bR\,\psi^{\dag}\psi \nonumber \\ \label{Eq:EffLagrangian}
\end{eqnarray}
and the fields $\psi(\br,\tau)=[\psi_{\uparrow}(\br,\tau),\psi_{\downarrow}(\br,\tau)]^T$
are eight-component Grassmann spinors in layer (1 and 2), valley ($\pm\bK$), and spin ($\uparrow$
and $\downarrow$) spaces and
\begin{equation}
\psi_{\sigma}(\br,\tau)=
\begin{bmatrix}
b_{1,\bK,\sigma}(\br,\tau) \\
b_{2,\bK,\sigma}(\br,\tau) \\
b_{1,-\bK,\sigma}(\br,\tau) \\
b_{2,-\bK,\sigma}(\br,\tau)
\end{bmatrix}.
\end{equation}
We omit the explicit dependence of these fields on $\br$ and $\tau$ for brevity; all
of these fields are at the same point and imaginary time.  The matrix $H(\bp)$ is
\begin{eqnarray}
H(\bp)&=&\frac{1}{2m^*}\left [(p^2_x-p^2_y)\Sigma^x+2p_xp_y\Sigma^y\right ] \cr
&+&v_3\left(p_x\Lambda^x+p_y\Lambda^y\right )+\Delta_{\text{app}}1\sigma^z1,
\end{eqnarray}
and the $\Sigma$ and $\Lambda$ matrices are
\begin{eqnarray}
\Sigma^x&=&1\sigma^x 1,\;\;\Sigma^y=\tau^z\sigma^y 1 \\
\Lambda^x&=&\tau^z\sigma^x 1,\;\;\Lambda^y=-1\sigma^y 1.
\end{eqnarray}
In deriving this low-energy theory, one also obtains an additional term, proportional
to $p^2\frac{\partial}{\partial\tau}$, that gives the dispersion the well-known ``Mexican hat''
shape.  However, it turns out that, under the RG transformation that was employed in
the previous work\cite{CvetkovicPRB2012, ThrockmortonPRB2012} and will be employed here,
the coefficient of this term is irrelevant in the RG sense; therefore, we do not include
it in our analysis.

The experimental values\cite{Mayorov2011,VelascoNatNano2012} of the effective mass $\ms$
and of the trigonal warping velocity $v_3$ are $\ms\approx 0.028m_e$ and $v_3\approx 1.41\times 10^5\text{ m/s}$.
The above low-energy theory is valid up to energy scales of $200\text{ meV}$, since
it is at this energy that the high-energy bands begin to affect the physics of the
system, and thus where the low-energy theory breaks down.  In fact, for the purpose
of our later renormalization group calculations, we will impose a momentum cutoff
$\Lambda$ corresponding to this energy scale, i.e. $\frac{\Lambda^2}{2\ms}=200\text{ meV}$.

In the last term, the interaction term, $S$ is the set of $16$ matrices operating on
valley and layer space.  In the zero-field case, there are $9$ independent coupling
constants\cite{CvetkovicPRB2012,ThrockmortonPRB2012,VafekPRB2010,LemonikPRB2010,LemonikPRB2012},
all multiplying these four-fermion interaction terms.  These independent coupling
constants correspond to representations of the $D_{3d}$ point symmetry group of the
Hamiltonian at the $\Gamma$ point in the absence of an electric field and to representations
of the $D_3$ symmetry group at the parabolic touching points, $\pm\bK$.  Application
of an electric field reduces the point symmetry group of the lattice and at the $\Gamma$
point to $C_{3v}$, but leaves all other symmetries (translation, spin $SU(2)$, and time
reversal) intact.  Similarly, the point symmetry group at $\pm\bK$ is reduced to $C_3$.
This would, in general, allow for additional four-fermion interactions and, as a consequence,
more coupling constants.  These possible interaction terms are
\begin{eqnarray}
&&g_{A_1,3,c}\int d^2{\br}\,(\psi^{\dag}\psi)(\psi^{\dag}1\sigma^z\psi) \cr
&&+g_{A_2,3,c}\int d^2{\br}\,(\psi^{\dag}\tau^z\sigma^z\psi)(\psi^{\dag}\tau^z 1\psi) \cr
&&+g_{A_1,3,s}\int d^2{\br}\,(\psi^{\dag}\vec{\sigma}\psi)\cdot(\psi^{\dag}1\sigma^z\vec{\sigma}\psi) \cr
&&+g_{A_2,3,s}\int d^2{\br}\,(\psi^{\dag}\tau^z\sigma^z\vec{\sigma}\psi)\cdot(\psi^{\dag}\tau^z 1\vec{\sigma}\psi).
\end{eqnarray}
These are the possible coupling constants before applying Fierz identities, which are
\begin{eqnarray}
(\psi^{\dag}S\psi)(\psi^{\dag}T\psi)=-\tfrac{1}{64}\sum_{a,b}\mbox{Tr}(S\Lambda_b T\Lambda_a)(\psi^{\dag}\Lambda_a\psi)(\psi^{\dag}\Lambda_b\psi), \nonumber \\
\end{eqnarray}
where $\Lambda_a$ and $\Lambda_b$ range over all matrices of the form $\tau^i\sigma^j s^k$,
and $\tau$ acts in the valley subspace, $\sigma$ in the layer subspace, and $s$ in
the spin subspace.  All fields in the above identity are at the same position and
imaginary time.  Some of these interaction terms would be eliminated if we were
to apply these identities.

However, these coupling constants will always be zero in any realistic situation because
they will not be generated under our RG transformation unless they start with non-zero
values.  They also do not generate, nor are generated by, the terms that are possible in
the zero-field case via the above Fierz identities.  We may see this by noting that the
additional terms only allowed in the non-zero-field case are all of the form,
\begin{equation}
g_{ST}\int d^2{\br}\,(\psi^{\dag}S\psi)(\psi^{\dag}T\psi),
\end{equation}
where $S\neq T$, while those that appear in the zero-field case all have $S=T$.  For these
reasons, we will not include these additional interaction terms in our analysis.  This
restriction to nine coupling constants, and no more, keeps our RG analysis reasonably tractable.

Note that we also include a chemical-potential-like term proportional to $\mu'$.  This
is a counterterm that is introduced in order to cancel out similar terms that are
generated from the interaction terms.  While this term must, strictly speaking, be
present, we do not explicitly calculate this term for similar reasons as those
discussed in the previous work\cite{ThrockmortonPRB2012}, namely, that it would be
cumbersome and unnecessary to do so.

The character tables of $C_{3v}$ and $C_3$ are as follows\cite{TinkhamBook}:
\begin{center}
    \begin{tabular}{ | l | l | l | l |}
    \hline
    ${\bf C_{3v}}$ & $E$ & 2$C_3$ & 3$\sigma_v$ \\
    \hline\hline
    $A_1$          & 1   & 1      & 1 \\ \hline
    $A_2$          & 1   & 1      & -1 \\ \hline
    $E$            & 2   & -1     & 0 \\ \hline
    \end{tabular}
\end{center}
\begin{center}
    \begin{tabular}{ | l | l | l | l |}
    \hline
    ${\bf C_{3}}$ & $E$ & $C_3$      & $(C_3)^2$ \\
    \hline\hline
    $A$           & 1   & 1          & 1  \\ \hline
    $E$ $(1)$     & 1   & $\omega$   & $\omega^2$ \\ \hline
    $E$ $(2)$     & 1   & $\omega^2$ & $\omega$  \\ \hline
    \end{tabular}
\end{center}
In the second table, $\omega=e^{2\pi i/3}$.

The classification of the $16$ $4\times 4$ matrices acting in valley and layer space
with respect to their transformation properties under the symmetries of our system are
\begin{eqnarray}
A_{1}+&:& 1_4,1\sigma^z \nonumber \\
A_{2}-&:&\tau^z\sigma^z,\tau^z1 \nonumber \\
E+&:&(1\sigma^x,\tau^z\sigma^y) \nonumber \\
E-&:& (\tau^z\sigma^x,-1\sigma^y) \nonumber \\
A_{\bK}+&:&(\tau^x\sigma^x;\tau^y\sigma^x) \nonumber \\
A_{\bK}-&:&(\tau^x\sigma^y;\tau^y\sigma^y) \nonumber \\
E_{\bK}+&:&(\tau^x 1,-\tau^y\sigma^z;-\tau^y 1,-\tau^x\sigma^z). \nonumber
\end{eqnarray}
The $\pm$ in front of the representation name denotes whether the particular operator is even or odd
under time reversal.  Groups of matrices enclosed in parentheses can be ``rotated'' into one another
by the symmetries of the system; matrices separated by commas are ``rotated'' into each other by the
point operations, while groups separated by semicolons are ``rotated'' into each other by translations.
Note that some of the operators that transform identically under the  {\it geometric} symmetries of the
system transform differently under time reversal, and thus belong to different representations of the
full symmetry group of the system.  Also note that, even though $C_3$ has only one-dimensional irreducible
representations, the point operations can ``rotate'' a given matrix in the $E_{\bK}$ set into another;
this is because they form a {\it reducible}, two-dimensional, representation of $C_3$.

As a result of the reduced symmetry of our problem compared to the zero-field case, we
will change the notation for the coupling constants from what was used before\cite{CvetkovicPRB2012,ThrockmortonPRB2012}.
The new names for the coupling constants (left), along with their original names (right),
are $g_{A_1,1}=g_{A_{1g}}$, $g_{A_1,2}=g_{A_{2u}}$, $g_{A_2,1}=g_{A_{2g}}$, $g_{A_2,2}=g_{A_{1u}}$,
$g_{E+}=g_{E_g}$, $g_{E-}=g_{E_u}$, $g_{A_{K}+}=g_{A_{1K}}$, and $g_{A_{K}-}=g_{A_{2K}}$.
The coupling constant, $g_{E_{K}}$, is unchanged.

The relationships between a microscopic density-density interaction and the coupling constants in
our effective low-energy theory\cite{ThrockmortonPRB2012}, which we find to first order in the microscopic
interaction, are unchanged from the zero-field case; we repeat them here in our notation:
\begin{eqnarray}
g_{A_1,1}&=&\tfrac{1}{2}(V_{\|,0}+V_{\bot,N})A_{uc}, \label{Eq:gA1g_V} \\
g_{A_1,2}&=&\tfrac{1}{2}(V_{\|,0}-V_{\bot,N})A_{uc}, \\
g_{E_K}&=&\tfrac{1}{4}V_{\|,2K}A_{uc}, \label{Eq:gEK_V}
\end{eqnarray}
where $A_{uc}=\frac{3\sqrt{3}}{2}a^2$ is the area of a unit cell of the
lattice, and
\begin{eqnarray}
V_{\|,0}&=&\sum_{\bR}V_{\|}(\bR), \label{Eq:V_def_1} \\
V_{\bot,N}&=&\tfrac{1}{3}\sum_{\bR,\delta}V_{\bot}(\bR-\delta), \label{Eq:V_def_2} \\
V_{\|,2K}&=&\sum_{\bR}V_{\|}(\bR)\cos(2\bK\cdot\bR). \label{Eq:V_def_3}
\end{eqnarray}
As pointed out earlier, this means that, for any microscopic density-density interaction term,
any additional coupling constants that would have been possible due to the reduced symmetry
of the system compared to the zero-field case are still zero and would not affect the theory
at all.

\section{Renormalization group analysis}
\subsection{RG equations}
We employ a finite-temperature Wilson momentum shell renormalization group procedure\cite{CvetkovicPRB2012,ThrockmortonPRB2012,ChakravartyPRB1988,MillisPRB1993}.
To summarize, we integrate out electronic modes in thin shells in momentum space, and then rescale
the momenta and temperature to restore the theory to its previous form, but with rescaled constants.
This allows us to derive differential equations, which we will call flow equations, describing how
these constants evolve as we integrate out electronic modes.  The RG equations are the same as
the ones given in Ref.\ \onlinecite{CvetkovicPRB2012}, and we repeat them here for convenience
and completeness (adapted to the notation of the present paper):
\begin{equation}
\frac{dg_i}{d\ell}=\sum_{j,k}g_j g_k\sum_{a=1}^6 A_{ijk}^{(a)}\Phi_a[\nu_3(\ell),\delta_{\text{app}}(\ell),t(\ell)], \label{Eq:GFlow_EF}
\end{equation}
where the sums on $j$ and $k$ are over all nine (non-zero) independent coupling constants that appear
in our low-energy theory.  The dimensionless temperature $t$, trigonal warping velocity $\nu_3$, and
layer energy difference $\delta_{\text{app}}$ are defined as follows:
\begin{equation}
t=\frac{T}{\Lambda^2/2\ms},\,\nu_3=\frac{v_3}{\Lambda/2\ms},\,\delta_{\text{app}}=\frac{\Delta_{\text{app}}}{\Lambda^2/2\ms}
\end{equation}
The functions $\Phi_a$ are given by Eqs.\ \eqref{Eq:Phi_EField_1}--\eqref{Eq:Phi_EField_6} in
Appendix \ref{App:FuncDefs} and the coefficients $A_{ijk}^{(a)}$ are given by Eq.\ \eqref{Eq:ACoeffs}
in Appendix \ref{App:RGEquCoeffs}.  The temperature and trigonal warping velocity satisfy the
flow equations,
\begin{eqnarray}
\frac{dT}{d\ell}&=&2T\;\;\Rightarrow\;T(\ell)=T_0e^{2\ell}, \\
\frac{dv_3}{d\ell}&=&v_3\;\;\Rightarrow\;v_3(\ell)=v_{3,0}e^{\ell}.
\end{eqnarray}
On the other hand, the layer energy difference has a nontrivial rescaling behavior, given by
\begin{equation}
\frac{d\delta_{\text{app}}}{d\ell}=2\delta_{\text{app}}\left [1+F(\nu_3,\delta_{\text{app}},t)\sum_{i}b_i g_i\right ], \label{Eq:DAppFlow}
\end{equation}
where the function $F$ is given by Eq.\ \eqref{Eq:FFunc} in Appendix \ref{App:FuncDefs}
and the coefficients $b_i$ are given by Eq.\ \eqref{Eq:bCoeffs} in Appendix \ref{App:RGEquCoeffs}.

The equations for the coupling constants describe two competing tendencies.  The second-order
factors in the coupling constants, $g_j g_k$, tend to increase the absolute value of the coupling
constants, while the functions $\Phi_a$ tend to suppress this increase.  The net effect of
these tendencies is to cause the coupling constants to saturate at finite values as $\ell\to\infty$
at sufficiently high temperatures.  However, at a certain temperature, the critical temperature
$T_c$, the coupling constants diverge as $\ell\to\infty$; we associate this divergence with the
appearance of an instability toward one or more symmetry-breaking orders.  While the coupling
constants themselves diverge, {\it ratios} of these constants tend to finite values.  Unlike
the zero-field case, however, there is another parameter besides trigonal warping that tends
to suppress the increase of the absolute values of the coupling constants, namely the applied
layer energy difference.  A non-zero layer energy difference can lower $T_c$, or even drive it
to zero, thus eliminating any instabilities that would otherwise be present.

\subsection{Asymptotic behavior}
We now determine the asymptotic behavior of the coupling constants and the layer energy difference
at the critical temperature and as $\ell\to\infty$.  Before doing so, we require the asymptotic
behaviors of the $\Phi_a$ and $F$ functions.  This behavior depends on whether the layer energy
difference increases more quickly or more slowly than the temperature under our RG transformation.
We will assume that $\delta_{\text{app}}$ scales exponentially in the asymptotic limit:
\begin{equation}
\delta_{\text{app}}(\ell)\approx Ce^{(2+\eta_{\delta})\ell}
\end{equation}
Here, $\eta_{\delta}$ is the anomalous exponent acquired by the layer energy difference.  We will
see later that this assumed form does, in fact, satisfy the above RG equations in the asymptotic
limit and, in the process, derive a formula for $\eta_{\delta}$.

If $\eta_{\delta}<0$, then the asymptotic behaviors of the $\Phi_a$ functions are
\begin{eqnarray}
\Phi_{a}[\nu_3(\ell),\delta_{\text{app}}(\ell),t(\ell)]&=&\frac{e^{-2\ell}}{t_c}+\ldots\text{ for }a=1,2, \nonumber \\ \label{Eq:PhiAsymp12NegEta} \\
\Phi_{3}[\nu_3(\ell),\delta_{\text{app}}(\ell),t(\ell)]&=&-\frac{\nu_{3,0}^2}{12t_c^3}e^{-4\ell}+\ldots, \label{Eq:PhiAsymp3NegEta} \\
\Phi_{4}[\nu_3(\ell),\delta_{\text{app}}(\ell),t(\ell)]&=&\frac{\nu_{3,0}^2}{24t_c^3}e^{-4\ell}+\ldots, \label{Eq:PhiAsymp4NegEta} \\
\Phi_{5}[\nu_3(\ell),\delta_{\text{app}}(\ell),t(\ell)]&=&\frac{C^2}{24t_c^3}e^{-2(1-\eta_{\delta})\ell}+\ldots, \label{Eq:PhiAsymp5NegEta} \\
\Phi_{6}[\nu_3(\ell),\delta_{\text{app}}(\ell),t(\ell)]&=&\frac{C^2}{12t_c^3}e^{-2(1-\eta_{\delta})\ell}+\ldots, \label{Eq:PhiAsymp6NegEta}
\end{eqnarray}
and that of $F$ is
\begin{equation}
F[\nu_3(\ell),\delta_{\text{app}}(\ell),t(\ell)]=\frac{e^{-2\ell}}{2t_c}+\ldots. \label{Eq:FAsympNegEta}
\end{equation}
Here, $t_c$ is the dimensionless critical temperature and $\nu_{3,0}$ is the initial (dimensionless)
trigonal warping velocity.  Note that these forms differ from those quoted in Ref.\ \onlinecite{CvetkovicPRB2012}.

If, on the other hand, $\eta_{\delta}>0$, then the $\Phi_a$ functions are approximated by
\begin{eqnarray}
\Phi_{a}[\nu_3(\ell),\delta_{\text{app}}(\ell),t(\ell)]&=&\frac{e^{-(2+\eta_{\delta})\ell}}{|C|}+\ldots\text{ for }a=1,2,5,6, \nonumber \\ \label{Eq:PhiAsymp1256PosEta} \\
\Phi_{3}[\nu_3(\ell),\delta_{\text{app}}(\ell),t(\ell)]&=&-\frac{\nu_{3,0}^2}{|C|^3}e^{-(4+3\eta_{\delta})\ell}+\ldots, \label{Eq:PhiAsymp3PosEta} \\
\Phi_{4}[\nu_3(\ell),\delta_{\text{app}}(\ell),t(\ell)]&=&\frac{\nu_{3,0}^2}{|C|^3}e^{-(4+3\eta_{\delta})\ell}+\ldots, \label{Eq:PhiAsymp4PosEta}
\end{eqnarray}
and $F$ by
\begin{equation}
F[\nu_3(\ell),\delta_{\text{app}}(\ell),t(\ell)]=\frac{e^{-(2+\eta_{\delta})\ell}}{|C|}+\ldots. \label{Eq:FAsympPosEta}
\end{equation}

We will now determine the asymptotic behaviors of the coupling constants and of the layer energy
difference.  Let us begin with the case, $\eta_{\delta}<0$.  In this case, because the $\Phi_1$
and $\Phi_2$ terms in the flow equations dominate, the asymptotic forms of the equations, and thus
their solutions, are the same as in the zero-field case\cite{CvetkovicPRB2012}.  As stated
earlier, as the coupling constants themselves diverge, ratios of any two divergent constants tend
to finite values.  At the critical temperature and as $\ell\to\infty$, we find that we can ``collapse''
the nine flow equations for the coupling constants onto a single equation for one of the divergent
constants.  Let us choose one such constant, $g_{r}$, with respect to which we will find the coupling
constant ratios.  The flow equation for this constant becomes
\begin{equation}
\frac{dg_r}{d\ell}=\mathcal{A}_{(r)}g_r^2\frac{e^{-2\ell}}{2t_c},
\end{equation}
where
\begin{equation}
\mathcal{A}_{(r)}=\sum_{j,k=1}^9 \sum_{a=1}^2 A^{(a)}_{rjk}\rho_j^{(r)} \rho_k^{(r)}
\end{equation}
and
\begin{equation}
\rho_j^{(r)}=\left.\frac{g_j}{g_r}\right |_{t=t_c,\ell\to\infty}
\end{equation}
are the values that the coupling constant ratios approach as we integrate out all electronic modes.  All
of the other couplings can then be obtained by multiplying by the corresponding ratio.  Note that the above
definition of $\mathcal{A}_{(r)}$ is different from that given in Ref.\ \onlinecite{CvetkovicPRB2012} by
a factor of $2$.  The solution to the above equation is just
\begin{equation}
g_r(\ell)=\frac{2t_c}{\mathcal{A}_{(r)}}e^{2\ell}.
\end{equation}

We now make use of this result to show that the asymptotic behavior of $\delta_{\text{app}}$ is, in fact,
exponential.  Using Eq.\ \eqref{Eq:FAsympNegEta} and the behavior of $g_r(\ell)$, the flow equation
for $\delta_{\text{app}}$, Eq.\ \eqref{Eq:DAppFlow}, becomes, after substituting in the exponential
ansatz, $\delta_{\text{app}}(\ell)=Ce^{(2+\eta_{\delta})\ell}$,
\begin{equation}
(2+\eta_{\delta})e^{(2+\eta_{\delta})\ell}=2\left [1+\sum_i\frac{b_i\rho_i^{(r)}}{\mathcal{A}_{(r)}}\right ]e^{(2+\eta_{\delta})\ell}.
\end{equation}
We see that the equation is indeed satisfied, and in fact we can read off the value of $\eta_{\delta}$:
\begin{equation}
\eta_{\delta}=\frac{2}{\mathcal{A}_{(r)}}\sum_i b_i\rho_i^{(r)} \label{Eq:EtaDeltaNegEta}
\end{equation}

Now we consider the other case, $\eta_{\delta}>0$.  In this case, as can be seen from Eqs.\
\eqref{Eq:PhiAsymp1256PosEta}--\eqref{Eq:FAsympPosEta}, the $\Phi_1$, $\Phi_2$, $\Phi_5$, and $\Phi_6$
terms are dominant.  Once again, the nine coupling constant flow equations ``collapse'' onto a
single equation, but now they take the form,
\begin{equation}
\frac{dg_r}{d\ell}=\bar{\mathcal{A}}_{(r)}g_r^2\frac{e^{-(2+\eta_{\delta})\ell}}{|C|},
\end{equation}
where
\begin{equation}
\bar{\mathcal{A}}_{(r)}=\sum_{j,k=1}^9 \sum_{a=1,2,5,6} A^{(a)}_{rjk}\rho_j^{(r)} \rho_k^{(r)}.
\end{equation}
The solution of this equation is
\begin{equation}
g_r(\ell)=\frac{(2+\eta_{\delta})|C|}{\bar{\mathcal{A}}_{(r)}}e^{(2+\eta_{\delta})\ell}.
\end{equation}
We may show that the assumption of exponential asymptotic behavior for $\delta_{\text{app}}$ is
``self-consistent'' in the same way as before.  If we substitute the above form for $g_r$ into
the flow equation for $\delta_{\text{app}}$ along with the asymptotic form for $F$ given in Eq.\
\eqref{Eq:FAsympPosEta}, then we find that the exponential form, $\delta_{\text{app}}(\ell)=Ce^{(2+\eta_{\delta})\ell}$,
satisfies the equation, with $\eta_{\delta}$ given by
\begin{equation}
\eta_{\delta}=\frac{4\sum_i b_i\rho_i^{(r)}}{\bar{\mathcal{A}}_{(r)}-2\sum_i b_i\rho_i^{(r)}}. \label{Eq:EtaDeltaPosEta}
\end{equation}

\subsection{Free energy and susceptibilities}
In order to determine the dominant symmetry-breaking tendencies, we start by introducting source
terms into the action (or, equivalently the Lagrangian).  The additional terms in the Lagrangian
are
\begin{eqnarray}
\Delta L&=&\sum_{i=1}^{32}\Delta^{\text{ph}}_i\int d^2\br\,\psi^\dag O^{(i)}\psi \cr
&+&\tfrac{1}{2}\sum_{i=1}^{16}\Delta^{\text{pp}}_i\int d^2\br\,\psi^\dag\tilde{O}^{(i)}\psi^*+\text{c.c.}
\end{eqnarray}
Here, $O^{(i)}$ runs over all $8\times 8$ matrices acting on valley, layer, and spin space, while
$\tilde{O}^{(i)}$ runs only over the antisymmetric matrices.  Note that there are only $32$ terms
in the particle-hole (ph) term; here we have already accounted for spin SU(2) symmetry.  The
orders that these correspond to are listed in Table I of Ref.\ \onlinecite{CvetkovicPRB2012}.  While
one might expect that the presence of a layer energy difference will mix some of these orders
(for example, the layer-polarized state no longer breaks any symmetries, as the symmetries that
it would break are no longer present), it turns out that, to second order in the source terms,
there are no terms in the free energy that mix any of the listed orders, as we will see shortly.

Before determining the free energy, we need the RG flow equations for these source terms.  They
are similar to those already derived in Ref.\ \onlinecite{CvetkovicPRB2012}:
\begin{eqnarray}
\frac{d\ln\Delta^{\text{ph}}_i}{d\ell}&=&2+\sum_{j=1}^{9}\sum_{a=1}^6 B^{(a)}_{ij}g_j(\ell)\Phi_a\left[\nu_3(\ell),\delta_{\text{app}}(\ell),t(\ell)\right], \nonumber\\ \label{Eq:DLogDeltaPH} \\
\frac{d\ln\Delta^{\text{pp}}_i}{d\ell}&=&2+\sum_{j=1}^{9}\sum_{a=1}^6\tilde{B}^{(a)}_{ij}g_j(\ell)\Phi_a\left[\nu_3(\ell),\delta_{\text{app}}(\ell),t(\ell)\right], \nonumber\\ \label{Eq:DLogDeltaPP}
\end{eqnarray}
where the coefficients $B^{(a)}_{ij}$ and $\tilde{B}^{(a)}_{ij}$ are given by Eqs.\ \eqref{Eq:B121}--\eqref{Eq:Bt56}
in Appendix \ref{App:RGEquCoeffs}.  These equations can be easily integrated\cite{CvetkovicPRB2012} to
obtain explicit expressions for the source term coefficients in terms of the coupling constants:
\begin{equation}\label{Eq:DeltaIntegrated}
\Delta_i^{\text{ph/pp}}(\ell)=\Delta_i^{\text{ph/pp}}(0)e^{2\ell}\exp[\Omega_i^{\text{ph/pp}}(\ell)],
\end{equation}
where
\begin{eqnarray}
\Omega_i^{\text{ph}}(\ell)&=&\sum_{j=1}^{9}\sum_{a=1}^6 B^{(a)}_{ij}\int_{0}^{\ell}d\ell'\,g_j(\ell')\Phi_a(\nu_3,\delta_{\text{app}},t), \nonumber \\ \\
\Omega_i^{\text{pp}}(\ell)&=&\sum_{j=1}^{9}\sum_{a=1}^6 {\tilde B}^{(a)}_{ij}\int_{0}^{\ell}d\ell'\,g_j(\ell')\Phi_a(\nu_3,\delta_{\text{app}},t). \nonumber \\
\end{eqnarray}
Here, $\nu_3$, $\delta_{\text{app}}$, and $t$ are understood to be functions of $\ell'$.

As was the case with the flow equations for the four-fermion coupling constants, the asymptotic
forms for the equations for the source terms depend on whether we assume that the anomalous dimension
of the layer energy difference $\eta_{\delta}<0$ or $\eta_{\delta}>0$.  In the former case, the
forms obtained are exactly as in the zero-field case\cite{CvetkovicPRB2012}:
\begin{eqnarray}
\frac{d\ln\Delta^{\text{ph}}_i}{d\ell}&=&2+\eta_i^{\text{ph}}\;\;\mbox{as}\;\ell\rightarrow\infty, \label{Eq:DLogDeltaPHAsympNegEta} \\
\frac{d\ln\Delta^{\text{pp}}_i}{d\ell}&=&2+\eta_i^{\text{pp}}\;\;\mbox{as}\;\ell\rightarrow\infty, \label{Eq:DLogDeltaPPAsympNegEta}
\end{eqnarray}
where
\begin{equation}
\eta_i^{\text{ph/pp}}=\frac{2\mathcal{B}^{\text{ph/pp}}_{i(r)}}{\mathcal{A}_{(r)}}
\end{equation}
and
\begin{eqnarray}
\mathcal{B}^{\text{ph}}_{i(r)}&=&\sum_{j=1}^{9}\sum_{a=1}^2 B^{(a)}_{ij}\rho_j^{(r)} \cr
\mathcal{B}^{\text{pp}}_{i(r)}&=&\sum_{j=1}^{9}\sum_{a=1}^2 \tilde{B}^{(a)}_{ij}\rho_j^{(r)}.
\end{eqnarray}
Note again the difference by a factor of $2$ in this definition compared to that in Ref.\
\onlinecite{CvetkovicPRB2012}; this cancels the factor of $2$ difference in the definition
of $\mathcal{A}_{(r)}$, thus giving us the same result for $\eta_i^{\text{ph/pp}}$.

We arrive at the result for the case, $\eta_{\delta}>0$, in a similar fashion; the result
has the same form as before, but now
\begin{equation}
\eta_i^{\text{ph/pp}}=\frac{(2+\eta_{\delta})\bar{\mathcal{B}}^{\text{ph/pp}}_{i(r)}}{\bar{\mathcal{A}}_{(r)}},
\end{equation}
where
\begin{eqnarray}
\bar{\mathcal{B}}^{\text{ph}}_{i(r)}&=&\sum_{j=1}^{9}\sum_{a=1,2,5,6} B^{(a)}_{ij}\rho_j^{(r)}, \\
\bar{\mathcal{B}}^{\text{pp}}_{i(r)}&=&\sum_{j=1}^{9}\sum_{a=1,2,5,6} \tilde{B}^{(a)}_{ij}\rho_j^{(r)}.
\end{eqnarray}

Now that we have determined the RG flows of the source terms, we turn our attention to the
correction to the free energy due to these terms\cite{CvetkovicPRB2012,NelsonPRB1975}.  We
find that the contribution to the free energy per unit area $\delta f(\Delta)$ from the
source terms at second order in said terms is
\begin{eqnarray}
&&\delta f(\Delta)=  \label{Eq:FreeEnergy} \\
&&=-\frac{\ms}{16\pi}\sum_{i=1}^{32}\int_{0}^{\infty}d\ell\,e^{-4\ell}[\Delta_{i}^{\text{ph}}(\ell)]^2\sum_{a=1}^{6}\alpha_{a,i}^{\text{ph}}\Phi_{a}(\nu_3,\delta_{\text{app}},t) \nonumber \\
&&-\frac{\ms}{16\pi}\sum_{i=1}^{16}\int_{0}^{\infty}d\ell\,e^{-4\ell}|\Delta_{i}^{\text{pp}}(\ell)|^2\sum_{a=1}^{6}\alpha_{a,i}^{\text{pp}}\Phi_{a}(\nu_3,\delta_{\text{app}},t). \nonumber
\end{eqnarray}
We again suppress the explicit dependence of $\nu_3$, $\delta_{\text{app}}$, and $t$ for
brevity of notation.  The $\alpha$ coefficients are given in Appendix \ref{App:FreeEnergyCoeffs}
by Eqs.\ \eqref{Eq:Alpha12PH}--\eqref{Eq:Alpha56PP}.  From this result, we can determine
the susceptibilities by simply taking the second derivatives with respect to the initial
values of the source terms:
\begin{eqnarray}
\chi^{\text{ph}}_i&=&-\left.\frac{\partial^2 f}{\partial[\Delta_i^{\text{ph}}(\ell=0)]^2}\right |_{\Delta(\ell=0)=0}, \\
\chi^{\text{pp}}_i&=&-\left.\frac{\partial^2 f}{\partial[\mbox{Re}\,\Delta_i^{\text{pp}}(\ell=0)]^2}\right |_{\Delta(\ell=0)=0} \cr
&=&-\left.\frac{\partial^2 f}{\partial[\mbox{Im}\,\Delta_i^{\text{pp}}(\ell=0)]^2}\right |_{\Delta(\ell=0)=0}. \nonumber \\
\end{eqnarray}
In these equations, the subscript, $\Delta(\ell=0)=0$, means that the value of the second
derivative is to be evaluated for all source terms set to zero.

With these results, we may now find the behavior of the susceptibilities just above the
critical temperature.  Once again, the result depends on whether $\eta_{\delta}$ is positive
or negative.  If it is negative, then we obtain the same result as in the zero-field case\cite{CvetkovicPRB2012}.
The susceptibilities, as functions of temperature, are, just above the critical temperature,
\begin{equation}
\chi_i^{\text{ph/pp}}\propto (t-t_c)^{-\gamma_i^{\text{ph/pp}}},
\end{equation}
where the exponent $\gamma_i^{\text{ph/pp}}$ is given by
\begin{equation}
\gamma_i^{\text{ph/pp}}=\eta_i^{\text{ph/pp}}-1.
\end{equation}
Therefore, if $\eta_i^{\text{ph/pp}}>1$, then the susceptibility diverges and therefore the
system is unstable to the associated order parameter.

In the case that $\eta_{\delta}$ is positive, then, following a similar analysis, we obtain
the same form for the susceptibility, but now the exponent is given by
\begin{equation}
\gamma_i^{\text{ph/pp}}=\frac{2}{2+\eta_{\delta}}\eta_i^{\text{ph/pp}}-1.
\end{equation}
The condition for divergence of a given susceptibility is thus $\eta_i^{\text{ph/pp}}>1+\tfrac{1}{2}\eta_{\delta}$.

\subsection{Effect of an applied layer energy difference on the possible fixed rays}
Due to the differences in the asymptotic behavior of the $\Phi_a$ functions, depending on whether the
anomalous dimension of the layer energy difference, $\eta_{\delta}$, is positive or negative, the
equations giving the possible fixed rays will differ in each of these two cases.

In the case that $\eta_{\delta}$ is negative, the equations for the fixed rays, and thus their
solutions, are the same as those found in the zero-field case\cite{CvetkovicPRB2012} (modified
to account for the differences in the asymptotic forms of the $\Phi_a$ and in the definition
of $\mathcal{A}_{(r)}$).  The differential equation satisfied by a coupling constant ratio
$\rho_j^{(r)}=\frac{g_j}{g_r}$ (this time at finite $\ell$) in the asymptotic limit is
\begin{eqnarray}
\dot{\rho}_j^{(r)}=\frac{d\rho_j^{(r)}}{d\ell}=\frac{4t_c}{{\mathcal A}_{(r)}}\sum_{k,l}\rho_k^{(r)}\rho_l^{(r)}\sum_{a=1}^2 \left (A_{jkl}^{(a)}-A_{rkl}^{(a)}\rho_j\right ). \nonumber\\ \label{Eq:DotRhoNegEta}
\end{eqnarray}
If we then set the right-hand side of this equation to zero and solve, we obtain the possible
fixed rays for our system.  We then determine which fixed rays are stable using the standard
techniques.  If a fixed ray is stable, then, if the coupling constants start with values
sufficiently close to said fixed ray, then they will tend toward that ray.  The full set of
stable fixed rays for this case is given in Ref.\ \onlinecite{CvetkovicPRB2012}.

\begin{figure*}[ht]
\centering
\includegraphics[width=0.9\textwidth]{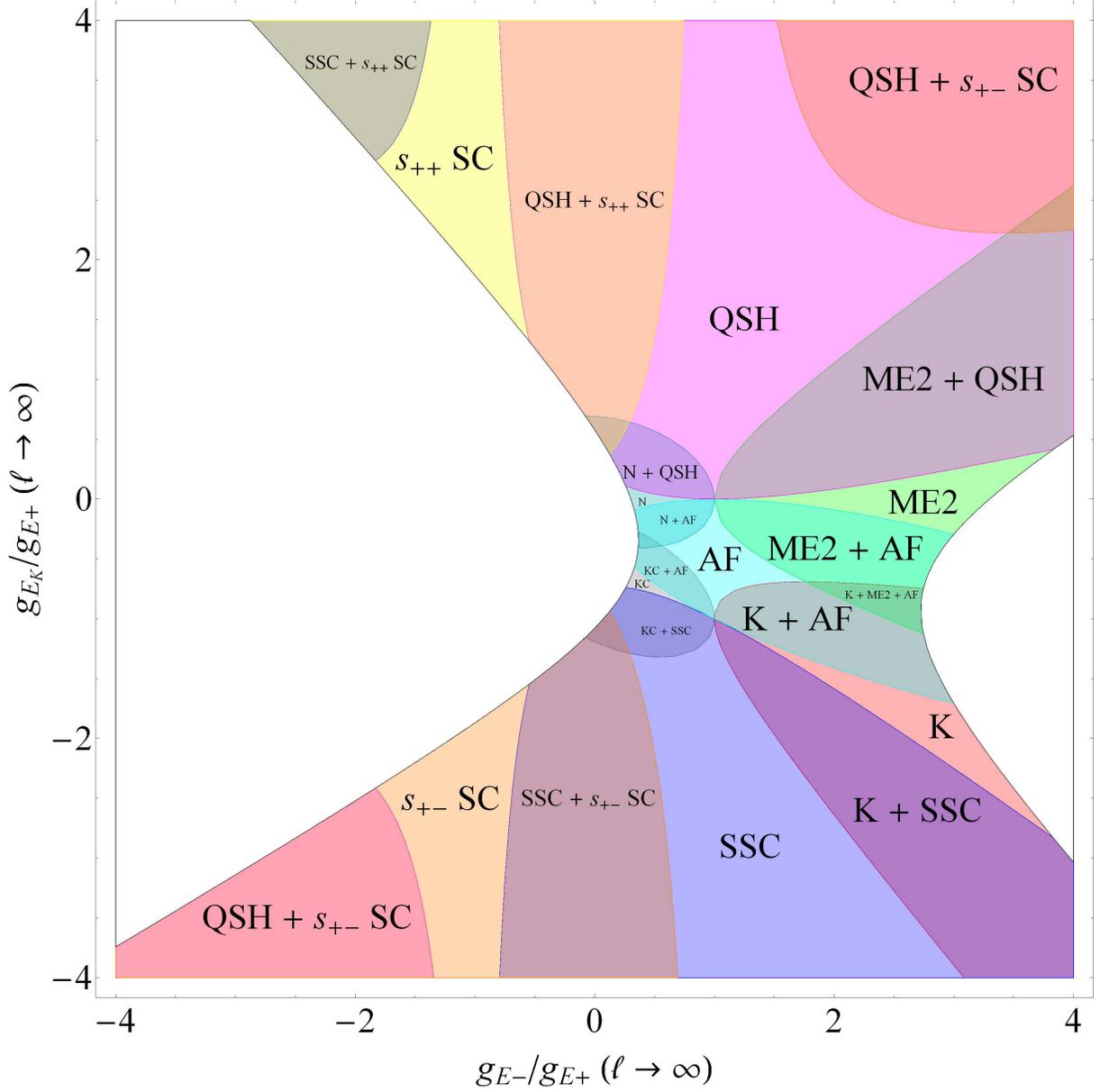}
\caption{\label{Fig:TargetPlane_PD}A plot of all of the phases found in the fixed plane
described by Eqs.\ (70)--(72) of Ref.\ \onlinecite{CvetkovicPRB2012} that are still allowed
when an electric field is present by the condition that the anomalous exponent of the applied
layer energy difference $\eta_{\delta}$ must be negative.  The white region is the ``forbidden''
region, within which $\eta_{\delta}$ becomes positive, in violation of this assumption.  The
possible phases are antiferromagnetic (AF, $A_1,2$ spin), quantum spin Hall (QSH, $A_2,1$ spin),
staggered spin current (SSC, $A_2,2$ spin), nematic (N, $E+$ charge), spontaneous ``bowtie''
current, or magnetoelectric (ME2, $E-$ charge), Kekul\'e (K, $A_{K}+$ charge), Kekul\'e current
(KC, $A_K-$ charge), $s_{++}$ superconductor ($s_{++}$ SC, $A_1,1$ singlet), and $s_{+-}$
superconductor ($s_{+-}$ SC, $A_2,1$ singlet), all of which are described in Ref.\ \onlinecite{CvetkovicPRB2012}.
In addition to this fixed plane, we also find nine isolated fixed rays.  Four of these are
described in Ref.\ \onlinecite{CvetkovicPRB2012}, while the other five are described in
the text.}
\end{figure*}

However, note that the above equation only holds true under the assumption that $\eta_{\delta}$
is negative.  This means that the fixed rays obtained from it are only valid if they are consistent
with this assumption.  We therefore check the fixed rays given in Ref.\ \onlinecite{CvetkovicPRB2012}
using Eq.\ \eqref{Eq:EtaDeltaNegEta} to make sure that they do in fact give us a negative
value of $\eta_{\delta}$.  Let us first consider the two-parameter family of fixed rays, or the
``target plane''.  In our new notation, the ratios given for this set of fixed rays are with
respect to $g_{E+}$.  The value of $\mathcal{A}_{(E+)}$ is
\begin{equation}
\mathcal{A}_{(E+)}=-3\frac{3+2x+3x^2+4y+4xy+8y^2}{1+x+2y}\frac{\ms}{4\pi}, \label{Eq:FixedPlane_A}
\end{equation}
where
\begin{equation}
x=\left.\frac{g_{E-}}{g_{E+}}\right |_{t=t_c,\ell\to\infty}\text{ and }y=\left.\frac{g_{E_K}}{g_{E+}}\right |_{t=t_c,\ell\to\infty}.
\end{equation}
If we now substitute the full set of fixed rays, given by Eqs.\ (68)--(70) of Ref.\ \onlinecite{CvetkovicPRB2012},
into our Eq.\ \eqref{Eq:EtaDeltaNegEta}, we obtain
\begin{equation}
\eta_{\delta}=\frac{5-18x+5x^2-4y-4xy-8y^2}{3(3+2x+3x^2+4y+4xy+8y^2)}. \label{Eq:EtaDelta_FixedPlane}
\end{equation}
The denominator of this expression is positive definite, meaning that any sign change in $\eta_{\delta}$
must come from the numerator.  If this set of fixed ratios is to be valid for all values of $x$ and $y$,
then the numerator must be negative definite.  However, this is not the case; there are values of $x$
and $y$ for which it becomes positive.  The fixed rays that correspond to such values are therefore no
longer valid.  We show a plot of the ``target plane'' with the ``forbidden'' region excluded in Fig.\ \ref{Fig:TargetPlane_PD}.
All of the isolated fixed rays, given by Eqs. (72)--(75) of Ref.\ \onlinecite{CvetkovicPRB2012},
on the other hand, are still valid.

There are two facts to note about the target plane.  First, note that the entire region that would have
corresponded to an instability toward a layer-polarized (LP) state in the zero-field case\cite{CvetkovicPRB2012}
is in the forbidden region; this is not surprising, as this state no longer breaks any of the symmetries
of our system.  Second of all, note that the $\left.\frac{g_{E-}}{g_{E+}}\right |_{t=t_c,\ell\to\infty}=\left.\frac{g_{E_K}}{g_{E+}}\right |_{t=t_c,\ell\to\infty}=0$
point is in the forbidden region as well.  This is the fixed ray that the system in the absence of an
applied layer energy difference would have tended to in the case corresponding to an infinite-range
repulsive (Coulomb) interaction, i.e., when only $g_{A_1+}$ starts out non-zero (and positive).  Instead,
the system will tend to a different fixed ray, which we will state shortly.

While some of the fixed rays that are present in the zero-field case are no longer valid, there
are also new fixed rays that were not previously present.  To find these, let us now assume that
$\eta_{\delta}>0$.  In this case, the flow equations for the ratios are similar in structure to
those for the previous case:
\begin{eqnarray}
\dot{\rho}_j^{(r)}=\frac{d\rho_j^{(r)}}{d\ell}=\frac {4 t_c}{\bar{\mathcal{A}}_{(r)}}  \sum_{k,l} \rho_k^{(r)} \rho_l^{(r)} \sum_{a=1,2,5,6} \left ( A_{jkl}^{(a)} - A_{rkl}^{(a)} \rho_j \right ). \nonumber\\ \label{Eq:DotRhoPosEta}
\end{eqnarray}
As before, to find the fixed rays, we simply set the right-hand side to zero and solve for the ratios.

Unlike in the $\eta_{\delta}<0$ case, we are unable to find the full set of fixed rays in a closed analytic
form.  However, it is possible to find some of the (isolated) fixed rays.  Of the solutions to the equations
for the fixed rays that we find, only five are admissible (i.e., they give us a positive value of $\eta_{\delta}$)
and stable.  We now list these new fixed rays and what instability they correspond to.  All ratios in the
following are at $t=t_c$ and in the limit $\ell\to\infty$.

\begin {itemize}
  \item [$R'_1$:]
    \begin{eqnarray}
      \frac{g_{A_1,1}}{g_{E+}}&=&\frac{1}{\sqrt{15}},\nonumber\\
	  \frac{g_{A_1,2}}{g_{E+}}&=&\sqrt{\frac{5}{3}},\nonumber\\
      \frac{g_{A_2,1}}{g_{E+}}&=&\frac{g_{A_2,2}}{g_{E+}}=-\frac{g_{E_K}}{g_{E+}}=-\sqrt{\frac{3}{5}},\nonumber\\
	  \frac{g_{E-}}{g_{E+}}&=&-\frac{g_{A_K+}}{g_{E+}}=\frac{g_{A_K-}}{g_{E+}}=-1,
    \end{eqnarray}
    and $g_{E+}(\ell\to\infty)<0$.  In this case, the only divergent susceptibility is toward an $s_{++}$
	superconducting state.  In this state, an $s$-wave superconducting gap opens with the same sign in both
	layers.
  \item [$R'_2$:]
    \begin{eqnarray}
      \frac{g_{A_1,1}}{g_{E+}}&=&-\frac{1}{\sqrt{15}},\nonumber\\
	  \frac{g_{A_1,2}}{g_{E+}}&=&-\sqrt{\frac{5}{3}},\nonumber\\
      \frac{g_{A_2,1}}{g_{E+}}&=&\frac{g_{A_2,2}}{g_{E+}}=-\frac{g_{E_K}}{g_{E+}}=\sqrt{\frac{3}{5}},\nonumber\\
	  \frac{g_{E-}}{g_{E+}}&=&-\frac{g_{A_K+}}{g_{E+}}=\frac{g_{A_K-}}{g_{E+}}=-1,
    \end{eqnarray}
    and  $g_{E+}(\ell\to\infty)>0$.  In this case, only the susceptibility toward an $s_{+-}$ superconducting
	state diverges.  Like the $s_{++}$ case, an $s$-wave superconducting gap opens, but now with opposite signs
	in each layer.
  \item [$R'_3$:]
    \begin{eqnarray}
      \frac{g_{A_1,1}}{g_{E+}}&=&\frac{g_{A_2,2}}{g_{E+}}=\frac{g_{A_K+}}{g_{E+}}=\frac{g_{A_K-}}{g_{E+}}=\frac{g_{E_K}}{g_{E+}}=0,\nonumber\\
	  \frac{g_{A_2,1}}{g_{E+}}&=&-\frac{g_{A_1,2}}{g_{E+}}=1-\sqrt{\frac{5}{3}},\nonumber\\
      \frac{g_{E-}}{g_{E+}}&=&-4+\sqrt{15},
    \end{eqnarray}
    and $g_{E+}(\ell\to\infty)<0$.  This yields a nematic order\cite{CvetkovicPRB2012,ThrockmortonPRB2012,VafekYangPRB2010,LemonikPRB2010,LemonikPRB2012}
	($E_{+}$ charge).  This order breaks the rotational symmetry of the system.  It does not open a gap in the
	electronic spectrum, but it does reconstruct the spectrum within each valley so that two of the four Dirac
	cones become gapped.
  \item [$R'_4$:]
    \begin{eqnarray}
      \frac{g_{A_1,1}}{g_{A_2,2}}&=&\frac{g_{A_2,1}}{g_{A_2,2}}=\frac{g_{E+}}{g_{A_2,2}}=\frac{g_{E-}}{g_{A_2,2}}=\frac{g_{E_K}}{g_{A_2,2}}=0,\nonumber\\
	  \frac{g_{A_1,2}}{g_{A_2,2}}&=&-1,\nonumber\\
      \frac{g_{A_K+}}{g_{A_2,2}}&=&\tfrac{1}{2}(-3-\sqrt{15}),\nonumber\\
	  \frac{g_{A_K-}}{g_{A_2,2}}&=&\tfrac{1}{2}(-3+\sqrt{15}),
    \end{eqnarray}
    and $g_{A_2,2}(\ell\to\infty)>0$.  In this case, we see an instability towards a Kekul\'e phase\cite{HouPRL2007}
	($A_{K}+$ charge).  This phase breaks translational symmetry by enlarging the unit cell into a ``supercell''
	consisting of three regular unit cells.  In two of the unit cells, the tight-binding hopping is changed on
	alternating bonds, while in the third the hoppings are unchanged.  This phase opens a gap in the electronic spectrum.
  \item [$R'_5$:]
     \begin{eqnarray}
      \frac{g_{A_1,1}}{g_{A_2,2}}&=&\frac{g_{A_2,1}}{g_{A_2,2}}=\frac{g_{E+}}{g_{A_2,2}}=\frac{g_{E-}}{g_{A_2,2}}=\frac{g_{E_K}}{g_{A_2,2}}=0,\nonumber\\
	  \frac{g_{A_1,2}}{g_{A_2,2}}&=&-1,\nonumber\\
      \frac{g_{A_K+}}{g_{A_2,2}}&=&\tfrac{1}{2}(-3+\sqrt{15}),\nonumber\\
	  \frac{g_{A_K-}}{g_{A_2,2}}&=&\tfrac{1}{2}(-3-\sqrt{15}),
    \end{eqnarray}
    and $g_{A_2,2}(\ell\to\infty)>0$.  Here, we obtain an instability towards a Kekul\'e current phase
	($A_{K}-$ charge).  This phase breaks both translational and time-reversal symmetry.  Like the Kekul\'e phase, the
	unit cell is enlarged into a ``supercell,'' but now, rather than modifying the tight-binding
	hoppings in two of the regular unit cells, we see a current circulating around these cells, in
	the same direction on both.  This phase is also gapped.
\end {itemize}

We mentioned earlier that the fixed ray that we would have approached in the case of an infinite-range
repulsive interaction in the absence of an applied layer energy difference is no longer valid; in this
case, we instead approach the ray given by $R'_3$ above.  The system is thus still unstable to a nematic
order.

\begin{figure*}[!p]
  \centering
  \includegraphics[width=0.49\textwidth]{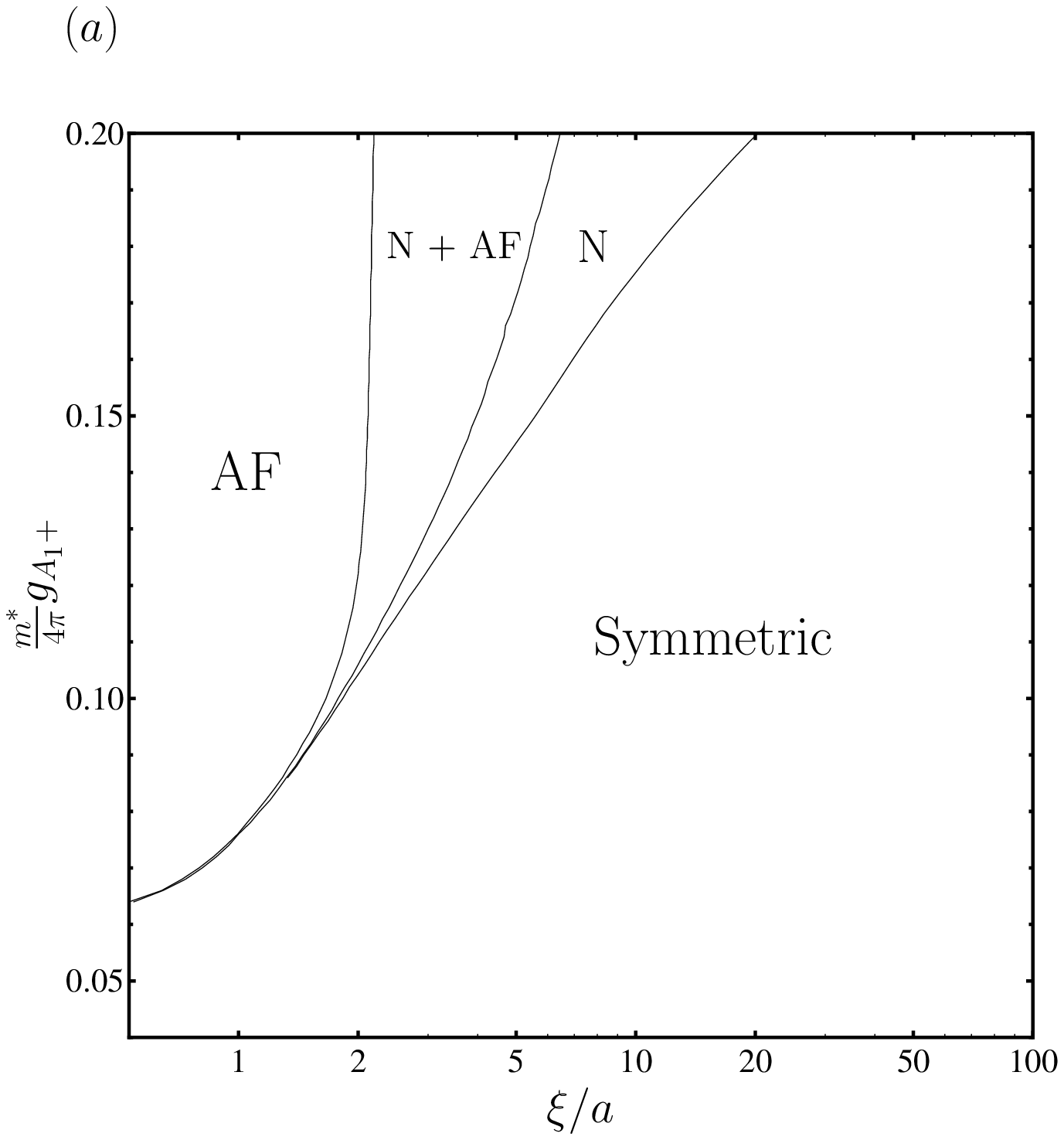}
  \includegraphics[width=0.49\textwidth]{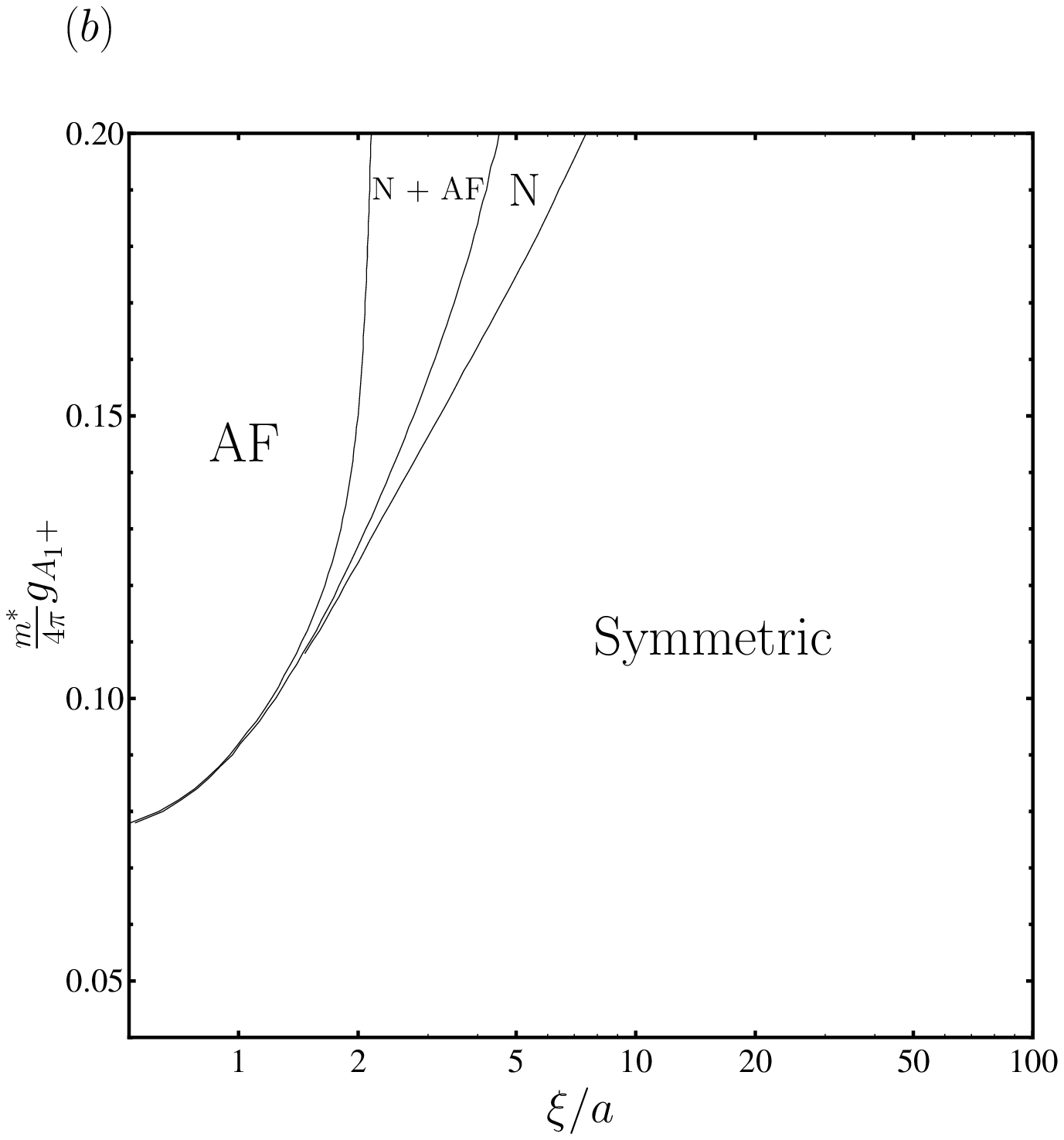}
  \includegraphics[width=0.49\textwidth]{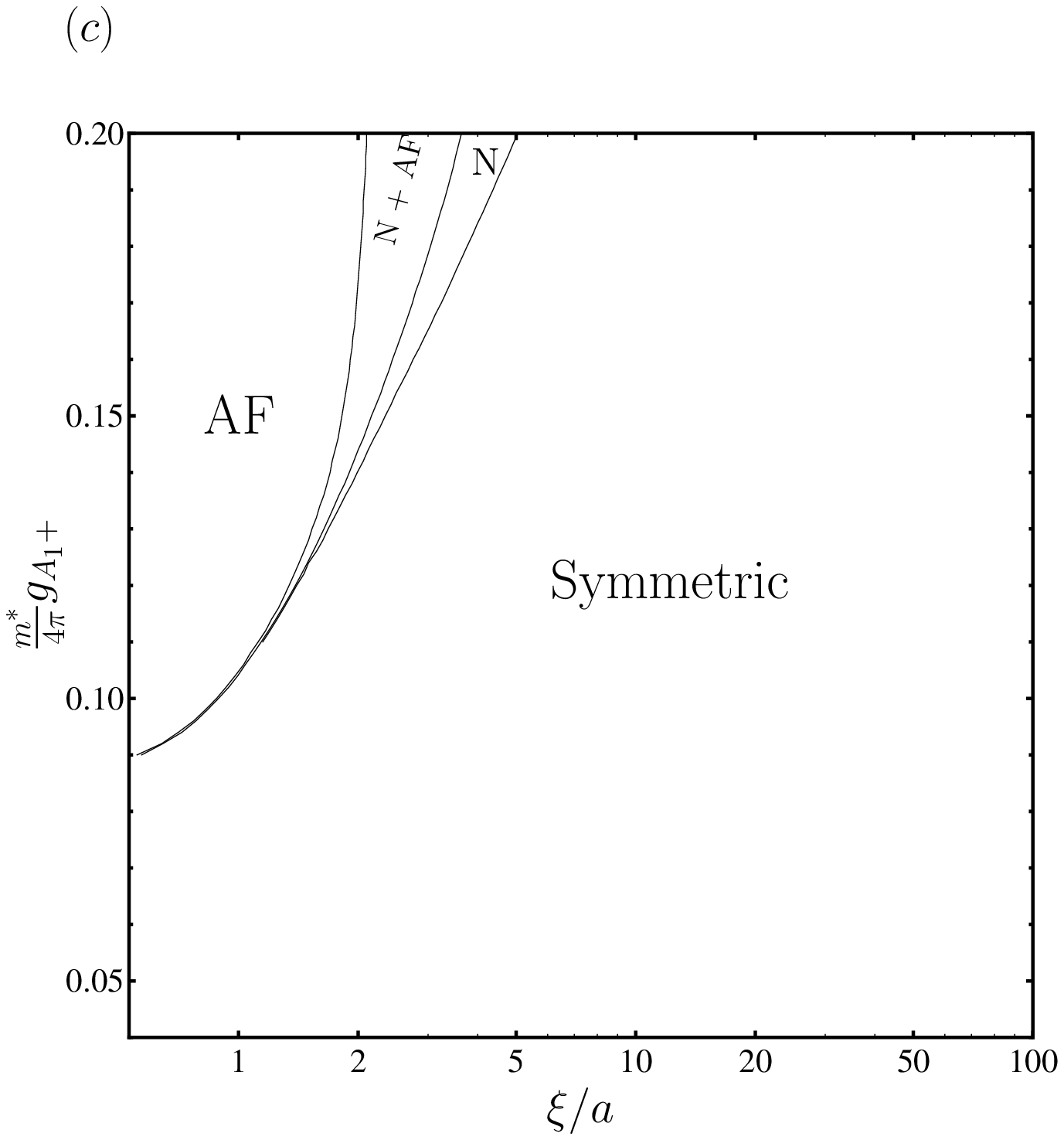}
  \includegraphics[width=0.49\textwidth]{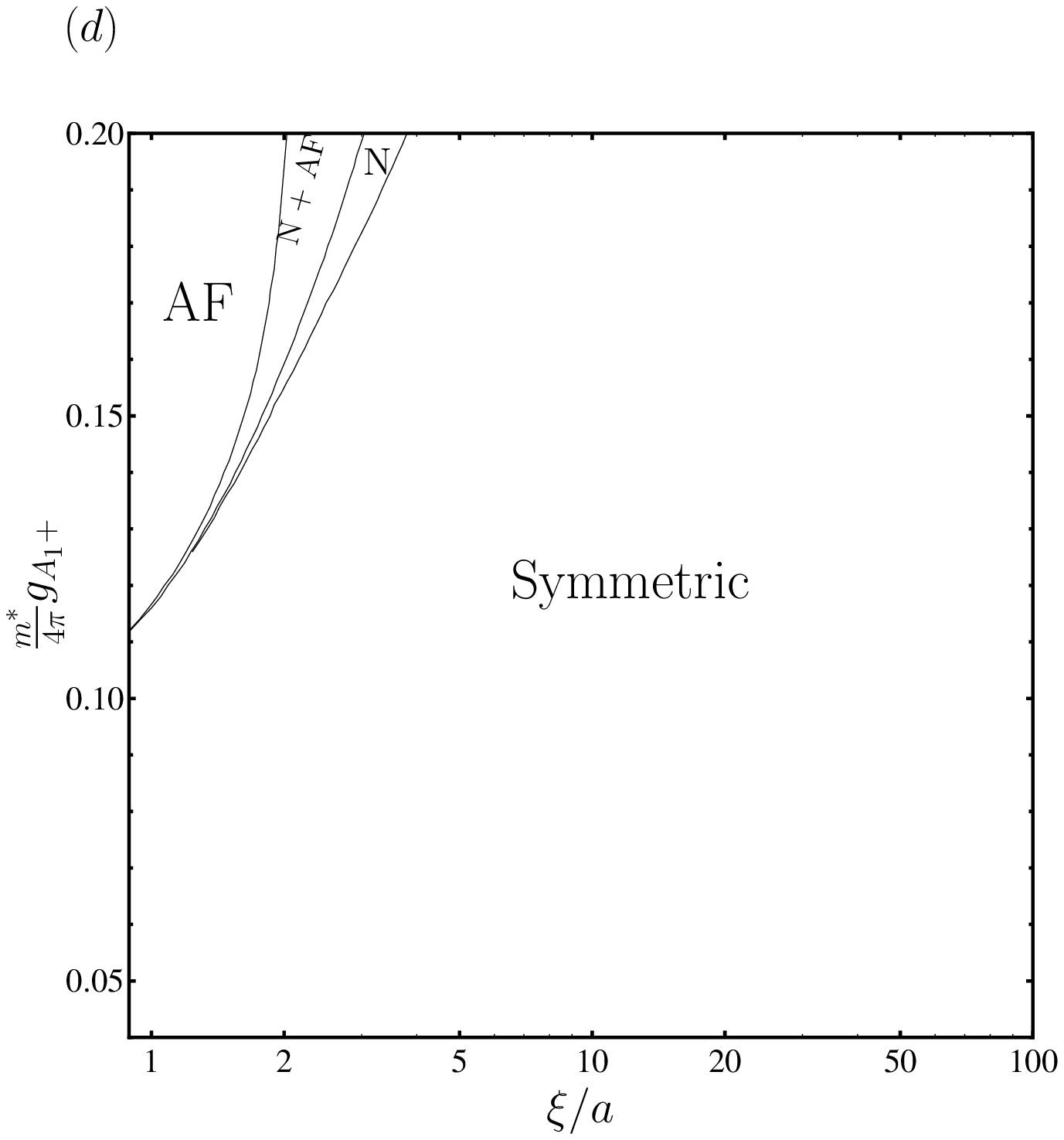}
  \caption{\label{Fig:PhaseMaps} Map of the leading instabilities for the interaction given by Eq.
  \eqref{Eq:IntScrCoul} as a function of overall interaction strength $\frac{\ms}{4\pi}g_{A_1+}$ and of the interaction
  range $\xi$ for a dimensionless applied layer energy difference (energy difference in units of $\Lambda^2/2\ms$)
  $\delta_{\text{app}}=0.036925$ $(a)$, $\delta_{\text{app}}=0.07385$ $(b)$, $\delta_{\text{app}}=0.110775$
  $(c)$, and $\delta_{\text{app}}=0.1477$ $(d)$.  In all four cases, we see the same pattern of phases
  that appears in the case with no applied layer energy difference\cite{ThrockmortonPRB2012}, namely that
  the system is unstable to an antiferromagnetic (AF) phase for short-ranged interactions, a nematic (N) phase
  for long-ranged interactions, and to both for intermediate ranges.  However, in the presence of an
  applied layer energy difference, the system has no instabilities for interactions that are too long-ranged
  or too weak.  The ranges at which the nematic instability first appears, at which the antiferromagnetic
  instability disappears, and at which the nematic instability disappears, all decrease with both increasing
  applied layer energy difference and with decreasing interaction strength.}
\end{figure*}

\section{Finite-range interaction}
We now turn our attention to the leading instability (or instabilities) of the system with finite-range
electron-electron interactions, as was done for the zero-field case in Ref.\ \onlinecite{ThrockmortonPRB2012}.
Once again, we will consider the Coulomb interaction screened by the presence of conducting plates parallel
to the sample.  In this work, however, we will only consider the case of the graphene sample exactly halfway
between two infinite conducting plates, as this corresponds to the only physical situation of the two considered
in the zero-field work in which we can simultaneously realize a finite electric field and maintain the sample at
half-filling.  The form of the microscopic electron-electron interaction is
\begin{equation}
V(\br)=U_0 \sum_{n=-\infty}^{\infty}\frac{(-1)^n}{\sqrt{(r/\xi)^2+n^2}}\approx U_0\frac{2\sqrt{2}e^{-\pi r/\xi}}{\sqrt{r/\xi}}. \label{Eq:IntScrCoul}
\end{equation}
Here, $U_0$ is the overall strength of the interaction, while $\xi$ is the distance between the conducting
plates and the sample, which sets the range of the interaction.  We may determine the corresponding starting
values of the coupling constants from Eqs.\ \eqref{Eq:gA1g_V}--\eqref{Eq:V_def_3}.  This interaction is, as
is, infinite for two particles on the same site.  We therefore set the interaction to a finite value of $1.2$
times the nearest-neighbor interaction.  The physical reason for doing this is that, in reality, two electrons
at the same site can no longer be treated as points for the purpose of determining how strongly they interact
due to the finite spread of their wave functions; this finite spread results in a finite on-site interaction.
The infinite interaction is simply an artifact of our model, and we regularize it by setting the interaction
to a large value.

In our calculation, we determine the leading instability as a function of the interaction range, parameterized
by $\xi$, and the interaction strength, parameterized by the initial value of $g_{A_1,1}$.  We do this
for four different values of the layer energy difference.  The values of $\delta_{\text{app}}$ that we
choose are $0.036925$, $0.07385$, $0.110775$ and $0.1477$.  In determining these maps of the leading
instability, we impose a lower cutoff on the critical temperature, $t_{c,\text{cutoff}}=10^{-10}$, below
which we consider the system to be in the symmetric phase (i.e.\ there are no instabilities).  In this
calculation, we set the trigonal warping velocity to zero; doing so simplifies the calculations, and we
do not expect this to change the overall qualitative picture that we obtain.  While we would not expect
any qualitative changes in the map even for the case with no electric field, we expect this statement to
hold even more strongly when there is a field present due to the fact that the electric field opens up a
gap in the electronic spectrum.  This is because the presence of this gap makes the effects of trigonal
warping less ``visible'' in the physics of the system.  In the absence of the field, it would change the
density of states noticeably, switching from that of a parabolic band structure to that of a Dirac cone
band structure at low energies.  When the field is applied, on the other hand, then no such distinct
qualitative change can easily be seen.

We show our results in Fig.\ \ref{Fig:PhaseMaps}.  As in the case with no layer energy difference\cite{ThrockmortonPRB2012},
we find that the system is unstable to an antiferromagnetic state for short interaction ranges, to a
nematic state for long ranges, and to both for intermediate ranges.  However, unlike in the zero-field
case, the critical ranges change noticeably with the interaction strength.  We also see that, below
a certain interaction strength or above a certain interaction range, the system has no instabilities
at all.  The critical interaction strength ranges from $\frac{\ms}{4\pi}g_{A_1+}\approx 0.011$ for
$\delta_{\text{app}}=0.1477$ to $\frac{\ms}{4\pi}g_{A_1+}\approx 0.064$ for $\delta_{\text{app}}=0.036925$,
implying a large change (by a factor of $6$) in the critical interaction strength for an approximate factor
of $4$ change in the layer energy difference.  The interaction range at which the system goes from a pure
antiferromagnetic instability to instabilities toward both the AF and nematic states remains steady at
around $\xi\approx 2.1a$ unless the interaction strength starts approaching the critical value.  The other
boundaries, namely those at which the antiferromagnetic and nematic instabilities disappear, on the other
hand, vary much more as the interaction strength is changed.  The interaction range at which the antiferromagnetic
instability disappears for an interaction strength of $\frac{\ms}{4\pi}g_{A_1+}=0.2$ ranges from $\xi\approx 3.04a$
for $\delta_{\text{app}}=0.1477$ to $\xi\approx 6.45a$ for $\delta_{\text{app}}=0.036925$, while the range
at which the nematic instability disappears for the same interaction strength varies from $\xi\approx 3.8a$
for $\delta_{\text{app}}=0.1477$ to $\xi\approx 20.29a$ for $\delta_{\text{app}}=0.036925$.  Thus, the main
quantitative effect of the applied electric field is a strong suppression of the tendency toward the instability
since a much larger critical interaction is necessary to induce a particular instability in general.

\section{Conclusion}
We have extended previous RG studies of the problem of interacting electrons in bilayer graphene at half
filling and at finite temperature\cite{CvetkovicPRB2012,ThrockmortonPRB2012} to include the effects of an
applied electric field perpendicular to the sample.  We consider both the problem of determining the possible
leading instabilities that one can in principle obtain and that of mapping out the leading instability of
the system for a Coulomb interaction screened by the presence of two parallel infinite conducting plates
as a function of the interaction strength and of the interaction range.

We discover that parts of the ``target plane'', a two-parameter family of fixed rays found in the zero-field
case\cite{CvetkovicPRB2012}, are no longer valid fixed rays, but that the four isolated fixed rays found therein
are still valid.  In particular, we find that the layer-polarized instability completely disappears; this is not surprising, as
such an instability no longer breaks any symmetries of the system.  However, a number of new stable fixed
rays appear, and we are able to determine five of them.  We then construct maps of the leading instability
(or instabilities) of the system as a function of interaction strength and interaction range for a screened
Coulomb-like interaction, like that between two electrons situated equidistant from two parallel infinite
conducting planes, for four values of the applied electric field, or, equivalently, of an applied layer
energy difference.   We also find that the instabilities of the system as a function of interaction strength
and range follow the same general pattern as they did in the absence of an electric field\cite{ThrockmortonPRB2012}.
We see an antiferromagnetic instability for short-ranged interactions, a nematic instability for long-ranged
interactions, and both instabilities for intermediate-ranged interactions.  However, these instabilities disappear
if the interaction range becomes too long or the interaction strength becomes too low.  Thus, the applied
electric field in general suppresses instabilities toward antiferromagnetic ground states.

Our results imply that an applied electric field can induce a nematic instability in a purely antiferromagnetic
sample, or eliminate any tendencies towards antiferromagnetism when tendencies toward both antiferromagnetic
and nematic instabilities are present in the absence of a field.  In the previous work at zero electric field,
one conclusion was that it was possible that, in some of the experiments, the system had both antiferromagnetic
and nematic orders\cite{ThrockmortonPRB2012}.  We predict that, should this in fact be the case, then, as one
applies an increasingly strong electric field, one should see the antiferromagnetic order disappear first, leaving
behind only a nematic order, followed by the nematic order vanishing as well for even stronger electric fields.
This may be seen by looking at Fig.\ \ref{Fig:PhaseMaps}.  Consider, for example, Fig.\ \ref{Fig:PhaseMaps}(a)
in particular, which is the leading instability map for the smallest layer energy difference considered, $\delta_{\text{app}}=0.036925$.
Note that, for an interaction range $\xi=5a$, that the system is unstable to both nematic and antiferromangetic orders
for the largest overall interaction strength considered, $\frac{\ms}{4\pi}g_{A_1+}=0.2$.  Compare this to Fig.\ \ref{Fig:PhaseMaps}(b),
which shows the same map, but now for a larger layer energy difference, $\delta_{\text{app}}=0.07385$---for the
same interaction range and interaction strength, the system is only unstable to a nematic phase.  If one now increases
the layer energy difference even further, then, as shown in Fig.\ \ref{Fig:PhaseMaps}(d), for which $\delta_{\text{app}}=0.1477$,
there are no instabilities at all for that interaction range and strength.

There are several mean field studies that have considered the effect of an electric field on the ground state
of bilayer graphene.  Two of these\cite{NandkishoreArXiv,GorbarPRB2012} conclude that the ground state should be
a quantum anomalous Hall state, in which the system has a non-zero Hall conductivity, even in the absence of an
applied magnetic field, while two others\cite{ZhangPRL2012,KharitonovPRB2012} conclude that it should be antiferromagnetic.
One other\cite{MinPRB2008} considers various ``(layer) pseudospin magnetic'' states.  In all cases, the application
of an electric field suppresses the predicted orders.  These works answer a different question than ours does---they
ask what the ground state of the system is, while we concern ourselves with the leading instability (or instabilities)
as we cool the system to the critical temperature at which symmetry-breaking order first appears.  The only overlap
between our RG study and these mean field studies is the prediction of suppression of symmetry-breaking order with
the application of a sufficiently strong electric field.  Note that most of these works do not consider the possibility
of a nematic state, and one\cite{NandkishoreArXiv} rules it out.  All of these only consider the possibility of a single
order parameter.  It is possible that, if these works had allowed for a coexistence of a nematic state with the gap-opening
states that they do consider, then they would have obtained a non-zero nematic order parameter as well.  While this
problem would be of interest, it is beyond the scope of this work.

The existence of an antiferromagnetic gap would be consistent with our prediction about short- and intermediate-ranged
interactions theoretically and with the results obtained in some experiments\cite{YacobyScience2010,VelascoNatNano2012}.
In these experiments, it is found that the observed gap decreases as an electric field is applied in either direction,
then increases again as the field is further strengthened.  This would be consistent with an antiferromagnetic order
being suppressed by the application of an electric field, thus reducing the overall gap, followed by an increase in
the gap entirely due to the action of the electric field.  While our theoretical predictions only consider the leading
instabilities of the system as we approach the critical temperature, we expect that the associated symmetry-breaking
orders should persist even well below said temperature.  In principle, one may see additional instabilities appear
as we cool the system below the critical temperature at which instabilities first appear, but investigating this interesting
possibility is beyond the scope of this work.  The fact that the cleanest bilayer systems report\cite{Mayorov2011}
only a nematic instability is also consistent with our finding, this time for long-ranged interactions, but much more
controlled experimental work as a function of an applied electric field will be necessary to verify our predictions in
the future.  Our main prediction is a general suppression of instabilities, but most particularly of antiferromagnetic
instability, in undoped bilayer graphene samples with an increasing externally applied electric field which creates a
layer energy difference.

\acknowledgements
This work is supported by LPS-CMTC and ARO-MURI.

\appendix

\section{Definition of functions appearing in RG equations} \label{App:FuncDefs}
We quote here the definitions of the $\Phi$ and $F$ functions appearing in Eqs.\ \eqref{Eq:GFlow_EF},
\eqref{Eq:DAppFlow}, \eqref{Eq:DLogDeltaPH}, and \eqref{Eq:DLogDeltaPP}.  The $\Phi$ functions are given by\cite{CvetkovicPRB2012}
\begin{eqnarray}
\Phi_1(\nu_3,\delta_{\text{app}},t)&=&\frac{1}{2\pi}\frac{1}{t}\int_{-1}^{1}\frac{dx}{\sqrt{1-x^2}}\Upsilon_1(x,\nu_3,\delta_{\text{app}},t), \label{Eq:Phi_EField_1} \\
\Phi_2(\nu_3,\delta_{\text{app}},t)&=&\frac{1}{\pi}\frac{1}{\nu_3}\int_{0}^{1}\frac{dx}{\sqrt{1-x^2}}\frac{1}{x}\Upsilon_2(x,\nu_3,\delta_{\text{app}},t), \\
\Phi_3(\nu_3,\delta_{\text{app}},t)&=&\frac{1}{\pi}\frac{1-\nu_3^2}{\nu_3}\int_{0}^{1}\frac{dx}{\sqrt{1-x^2}}\frac{1}{x}\Upsilon_3(x,\nu_3,\delta_{\text{app}},t), \nonumber\\ \\
\Phi_4(\nu_3,\delta_{\text{app}},t)&=&\frac{1}{2\pi}\frac{1}{t}\int_{-1}^{1}\frac{dx}{\sqrt{1-x^2}}\Upsilon_4(x,\nu_3,\delta_{\text{app}},t), \\
\Phi_5(\nu_3,\delta_{\text{app}},t)&=&\frac{1}{2\pi}\frac{\delta_{\text{app}}^2}{t}\int_{-1}^{1}\frac{dx}{\sqrt{1-x^2}}\Upsilon_5(x,\nu_3,\delta_{\text{app}},t), \nonumber\\ \\
\Phi_6(\nu_3,\delta_{\text{app}},t)&=&\frac{\delta_{\text{app}}^2}{1-\nu_3^2}\Phi_3(\nu_3,\delta_{\text{app}},t), \label{Eq:Phi_EField_6}
\end{eqnarray}
where the $\Upsilon$ functions are
\begin{eqnarray}
\Upsilon_1(x,\nu_3,\delta_{\text{app}},t)=\frac{2t}{Q_+}\tanh\left (\frac{Q_+}{2t}\right )+\frac{1}{\cosh^2\left (\frac{Q_+}{2t}\right )}, \nonumber \\
\end{eqnarray}
\begin{equation}
\Upsilon_2(x,\nu_3,\delta_{\text{app}},t)=\sum_{\lambda=\pm}\lambda Q_\lambda\tanh\left (\frac{Q_\lambda}{2t}\right ),
\end{equation}
\begin{equation}
\Upsilon_3(x,\nu_3,\delta_{\text{app}},t)=-\sum_{\lambda=\pm}\frac{\lambda}{Q_\lambda}\tanh\left (\frac{Q_\lambda}{2t}\right ),
\end{equation}
\begin{eqnarray}
&&\Upsilon_4(x,\nu_3,\delta_{\text{app}},t) \cr
&&=\left (\frac{Q_+^{(0)}}{Q_+}\right )^2\left [\frac{2t}{Q_+}\tanh\left (\frac{Q_+}{2t}\right )-\frac{1}{\cosh^2\left (\frac{Q_+}{2t}\right )}\right ], \nonumber \\
\end{eqnarray}
\begin{equation}
\Upsilon_5(x,\nu_3,\delta_{\text{app}},t)=\frac{1}{(Q_+^{(0)})^2}\Upsilon_4(x,\nu_3,\delta_{\text{app}},t),
\end{equation}
and
\begin{eqnarray}
Q_\pm&=&\sqrt{1+\nu_3^2+\delta_{\text{app}}^2\pm 2x\nu_3}, \\
Q_\pm^{(0)}&=&\sqrt{1+\nu_3^2\pm 2x\nu_3}. \label{Eq:Qpm_EField}
\end{eqnarray}
The $F$ function is given by
\begin{equation}
F(\nu_3,\delta_{\text{app}},t)=\frac{1}{\pi}\int_{-1}^{1}\frac{dx}{\sqrt{1-x^2}}\frac{1}{Q_+}\tanh\left (\frac{Q_+}{2t}\right ). \label{Eq:FFunc}
\end{equation}

\section{Coefficients of RG equations} \label{App:RGEquCoeffs}
Here, we quote the formulas for the $A$, $b$, and $B$ coefficients\cite{CvetkovicPRB2012} appearing in
Eqs.\ \eqref{Eq:GFlow_EF}, \eqref{Eq:DAppFlow}, \eqref{Eq:DLogDeltaPH}, and \eqref{Eq:DLogDeltaPP}.
The coefficients $A^{(a)}_{ijk}$ are given by
\begin{equation}
A_{ijk}^{(a)}=A_{ijk}^{(a)}(1)+A_{ijk}^{(a)}(2+3)+A_{ijk}^{(a)}(4)+A_{ijk}^{(a)}(5), \label{Eq:ACoeffs}
\end{equation}
where
\begin{widetext}
\begin{eqnarray}
A_{iii}^{(1/2)}(1)&=&-\tfrac{1}{2}\{8\pm\mbox{Tr}[(\Gamma_i^{(1)}\tau^z 1_4)^2]\}\frac{m^*}{4\pi}, \\
A_{iii}^{(3/4)}(1)&=&\tfrac{1}{4}\{\mbox{Tr}[(\Gamma_i^{(1)}1\sigma^x1)^2]\mp\mbox{Tr}[(\Gamma_i^{(1)}\tau^z\sigma^x1)^2]\mp\mbox{Tr}[(\Gamma_i^{(1)}1\sigma^y1)^2]+\mbox{Tr}[(\Gamma_i^{(1)}\tau^z\sigma^y1)^2]\}\frac{m^*}{4\pi},
\end{eqnarray}
\begin{eqnarray}
A_{iij}^{(1/2)}(2+3)&=&\tfrac{1}{8}\sum_{m=1}^{m_j}\{\mbox{Tr}[(\Gamma_i^{(1)}\Gamma_j^{(m)})^2]\pm\mbox{Tr}(\Gamma_i^{(1)}\Gamma_j^{(m)}\tau^z 1_4\Gamma_i^{(1)}\tau^z 1_4\Gamma_j^{(m)})\}\frac{m^*}{4\pi}, \\
A_{iij}^{(3/4)}(2+3)&=&-\tfrac{1}{16}\sum_{m=1}^{m_j}[\mbox{Tr}(\Gamma_i^{(1)}\Gamma_j^{(m)}1\sigma^x1\Gamma_i^{(1)}1\sigma^x1\Gamma_j^{(m)})\mp\mbox{Tr}(\Gamma_i^{(1)}\Gamma_j^{(m)}\tau^z\sigma^x1\Gamma_i^{(1)}\tau^z\sigma^x1\Gamma_j^{(m)}) \cr
&\mp&\mbox{Tr}(\Gamma_i^{(1)}\Gamma_j^{(m)}1\sigma^y1\Gamma_i^{(1)}1\sigma^y1\Gamma_j^{(m)})+\mbox{Tr}(\Gamma_i^{(1)}\Gamma_j^{(m)}\tau^z\sigma^y1\Gamma_i^{(1)}\tau^z\sigma^y1\Gamma_j^{(m)})]\frac{m^*}{4\pi},
\end{eqnarray}
\begin{eqnarray}
A_{kij}^{(1/2)}(4)&=&\tfrac{1}{128}\sum_{m=1}^{m_i}\sum_{n=1}^{m_j}[\mbox{Tr}(\Gamma_k^{(1)}\Gamma_i^{(m)}\Gamma_j^{(n)})\mbox{Tr}(\Gamma_k^{(1)}\Gamma_j^{(n)}\Gamma_i^{(m)})\pm\mbox{Tr}(\Gamma_k^{(1)}\Gamma_i^{(m)}\tau^z 1_4\Gamma_j^{(n)})\mbox{Tr}(\Gamma_k^{(1)}\Gamma_j^{(n)}\tau^z 1_4\Gamma_i^{(m)})]\frac{m^*}{4\pi}, \\
A_{kij}^{(3/4)}(4)&=&-\tfrac{1}{256}\sum_{m=1}^{m_i}\sum_{n=1}^{m_j}[\mbox{Tr}(\Gamma_k^{(1)}\Gamma_i^{(m)}1\sigma^x1\Gamma_j^{(n)})\mbox{Tr}(\Gamma_k^{(1)}\Gamma_j^{(n)}1\sigma^x1\Gamma_i^{(m)})\mp\mbox{Tr}(\Gamma_k^{(1)}\Gamma_i^{(m)}\tau^z\sigma^x1\Gamma_j^{(n)})\mbox{Tr}(\Gamma_k^{(1)}\Gamma_j^{(n)}\tau^z\sigma^x1\Gamma_i^{(m)}) \cr
&\mp&\mbox{Tr}(\Gamma_k^{(1)}\Gamma_i^{(m)}1\sigma^y1\Gamma_j^{(n)})\mbox{Tr}(\Gamma_k^{(1)}\Gamma_j^{(n)}1\sigma^y1\Gamma_i^{(m)})+\mbox{Tr}(\Gamma_k^{(1)}\Gamma_i^{(m)}\tau^z\sigma^y1\Gamma_j^{(n)})\mbox{Tr}(\Gamma_k^{(1)}\Gamma_j^{(n)}\tau^z\sigma^y1\Gamma_i^{(m)})]\frac{m^*}{4\pi}, \\
A_{kij}^{(5/6)}(4)&=&\tfrac{-1}{128}\sum_{m=1}^{m_i}\sum_{n=1}^{m_j}[\mbox{Tr}(\Gamma_k^{(1)}\Gamma_i^{(m)}1\sigma^z1\Gamma_j^{(n)})\mbox{Tr}(\Gamma_k^{(1)}\Gamma_j^{(n)}1\sigma^z1\Gamma_i^{(m)})\pm\mbox{Tr}(\Gamma_k^{(1)}\Gamma_i^{(m)}\tau^z\sigma^z1\Gamma_j^{(n)})\mbox{Tr}(\Gamma_k^{(1)}\Gamma_j^{(n)}\tau^z\sigma^z1\Gamma_i^{(m)})]\frac{m^*}{4\pi}, \nonumber \\
\end{eqnarray}
and
\begin{eqnarray}
A_{kij}^{(1/2)}(5)&=&-\tfrac{1}{128}\sum_{m=1}^{m_i}\sum_{n=1}^{m_j}\{[\mbox{Tr}(\Gamma_k^{(1)}\Gamma_i^{(m)}\Gamma_j^{(n)})]^2\mp[\mbox{Tr}(\Gamma_k^{(1)}\Gamma_i^{(m)}\tau^z 1_4\Gamma_j^{(n)})]^2\}\frac{m^*}{4\pi}, \\
A_{kij}^{(3/4)}(5)&=&-\tfrac{1}{256}\sum_{m=1}^{m_i}\sum_{n=1}^{m_j}\{[\mbox{Tr}(\Gamma_k^{(1)}\Gamma_i^{(m)}1\sigma^x1\Gamma_j^{(n)})]^2\pm[\mbox{Tr}(\Gamma_k^{(1)}\Gamma_i^{(m)}\tau^z\sigma^x1\Gamma_j^{(n)})]^2 \cr
&\pm&[\mbox{Tr}(\Gamma_k^{(1)}\Gamma_i^{(m)}1\sigma^y1\Gamma_j^{(n)})]^2+[\mbox{Tr}(\Gamma_k^{(1)}\Gamma_i^{(m)}\tau^z\sigma^y1\Gamma_j^{(n)})]^2\}\frac{m^*}{4\pi}, \\
A_{kij}^{(5/6)}(5)&=&-\tfrac{1}{128}\sum_{m=1}^{m_i}\sum_{n=1}^{m_j}\{[\mbox{Tr}(\Gamma_k^{(1)}\Gamma_i^{(m)}1\sigma^z1\Gamma_j^{(n)})]^2\pm[\mbox{Tr}(\Gamma_k^{(1)}\Gamma_i^{(m)}\tau^z\sigma^z1\Gamma_j^{(n)})]^2\}\frac{m^*}{4\pi}.
\end{eqnarray}
In these expressions, the top signs correspond to the first number in the superscript on the left-hand side, while the bottom corresponds to the second.  The $8\times 8$ matrices $\Gamma_i^{(m)}$ appearing in these expressions are defined as follows:
\begin{eqnarray}
\Gamma_1^{(1)}&=&1_8, \label{Eq:GammaMat1} \\
\Gamma_2^{(1)}&=&\tau^z\sigma^z1_2, \\
\Gamma_3^{(1)}&=&1_2\sigma^x1,\,\Gamma_3^{(2)}=\tau^z\sigma^y1_2, \\
\Gamma_4^{(1)}&=&\tau^z1_4, \\
\Gamma_5^{(1)}&=&1_2\sigma^z1_2, \\
\Gamma_6^{(1)}&=&\tau^z\sigma^x1_2,\,\Gamma_6^{(2)}=-1_2\sigma^y1_2, \\
\Gamma_7^{(1)}&=&\tau^x\sigma^x1_2,\,\Gamma_7^{(2)}=\tau^y\sigma^x1_2, \\
\Gamma_8^{(1)}&=&\tau^x\sigma^y1_2,\,\Gamma_8^{(2)}=\tau^y\sigma^y1_2, \\
\Gamma_9^{(1)}&=&\tau^x1_4,\,\Gamma_9^{(2)}=-\tau^y\sigma^z1_2,\,\Gamma_9^{(3)}=-\tau^y1_4,\,\Gamma_9^{(4)}=-\tau^x\sigma^z1_2. \label{Eq:GammaMat9}
\end{eqnarray}
The superscripts $(m)$ refer to the multiplicity of a given representation.

The coefficients $B^{(a)}_{ij}$ in Eq.\ \eqref{Eq:DLogDeltaPH} are
\begin{equation}
B^{(a)}_{ij}=B^{(a)}_{ij}(1)+B^{(a)}_{ij}(2),
\end{equation}
where
\begin{eqnarray}
B^{(1/2)}_{ij}(1)&=&-\tfrac{1}{2}\sum_{n=1}^{m_j}[\mbox{Tr}(O^{(i)}\Gamma_j^{(n)})\pm\mbox{Tr}(\tau^z 1_4O^{(i)}\tau^z 1_4\Gamma_j^{(n)})]\frac{m^*}{4\pi}, \label{Eq:B121} \\
B^{(3/4)}_{ij}(1)&=&\tfrac{1}{4}\sum_{n=1}^{m_j}[\mbox{Tr}(1\sigma^x1O^{(i)}1\sigma^x1\Gamma_j^{(n)})\mp\mbox{Tr}(\tau^z\sigma^x1O^{(i)}\tau^z\sigma^x1\Gamma_j^{(n)}) \cr
&\mp&\mbox{Tr}(1\sigma^y1O^{(i)}1\sigma^y1\Gamma_j^{(n)})+\mbox{Tr}(\tau^z\sigma^y1O^{(i)}\tau^z\sigma^y1\Gamma_j^{(n)})]\frac{m^*}{4\pi}, \\
B^{(5/6)}_{ij}(1)&=&\tfrac{1}{2}\sum_{n=1}^{m_j}[\mbox{Tr}(1\sigma^z1O^{(i)}1\sigma^z1\Gamma_j^{(n)})\pm\mbox{Tr}(\tau^z\sigma^z1O^{(i)}\tau^z\sigma^z1\Gamma_j^{(n)})], \\
B^{(1/2)}_{ij}(2)&=&\tfrac{1}{16}\sum_{n=1}^{m_j}\{\mbox{Tr}[(O^{(i)}\Gamma_j^{(n)})^2]\pm\mbox{Tr}(O^{(i)}\Gamma_j^{(n)}\tau^z 1_4O^{(i)}\tau^z 1_4\Gamma_j^{(n)})\}\frac{m^*}{4\pi}, \label{Eq:B122} \\
B^{(3/4)}_{ij}(2)&=&-\tfrac{1}{32}\sum_{n=1}^{m_j}[\mbox{Tr}(O^{(i)}\Gamma_j^{(n)}1\sigma^x1O^{(i)}1\sigma^x1\Gamma_j^{(n)})\mp\mbox{Tr}(O^{(i)}\Gamma_j^{(n)}\tau^z\sigma^x1O^{(i)}\tau^z\sigma^x1\Gamma_j^{(n)}) \cr
&\mp&\mbox{Tr}(O^{(i)}\Gamma_j^{(n)}1\sigma^y1O^{(i)}1\sigma^y1\Gamma_j^{(n)})+\mbox{Tr}(O^{(i)}\Gamma_j^{(n)}\tau^z\sigma^y1O^{(i)}\tau^z\sigma^y1\Gamma_j^{(n)})]\frac{m^*}{4\pi}, \\
B^{(5/6)}_{ij}(2)&=&-\tfrac{1}{16}\sum_{n=1}^{m_j}[\mbox{Tr}(O^{(i)}\Gamma_j^{(n)}1\sigma^z1O^{(i)}1\sigma^z1\Gamma_j^{(n)})\pm\mbox{Tr}(O^{(i)}\Gamma_j^{(n)}\tau^z\sigma^z1O^{(i)}\tau^z\sigma^z1\Gamma_j^{(n)})]\frac{m^*}{4\pi}. \label{Eq:B562}
\end{eqnarray}
The coefficients $\tilde{B}^{(a)}_{ij}$ in Eq.\ \eqref{Eq:DLogDeltaPP} are given by
\begin{eqnarray}
{\tilde B}^{(1/2)}_{ij}&=&-\tfrac{1}{16}\sum_{n=1}^{m_j}\{\mbox{Tr}[{\tilde O}^{(i)}\Gamma_j^{(n)}{\tilde O}^{(i)}(\Gamma_j^{(n)})^T]\mp\mbox{Tr}[{\tilde O}^{(i)}\Gamma_j^{(n)}\tau^z 1_4{\tilde O}^{(i)}\tau^z 1_4(\Gamma_j^{(n)})^T]\}\frac{m^*}{4\pi}, \label{Eq:Bt12} \\
{\tilde B}^{(3/4)}_{ij}&=&-\tfrac{1}{32}\sum_{n=1}^{m_j}\{\mbox{Tr}[{\tilde O}^{(i)}\Gamma_j^{(n)}1\sigma^x1{\tilde O}^{(i)}1\sigma^x1(\Gamma_j^{(n)})^T]\pm\mbox{Tr}[{\tilde O}^{(i)}\Gamma_j^{(n)}\tau^z\sigma^x1{\tilde O}^{(i)}\tau^z\sigma^x1(\Gamma_j^{(n)})^T] \cr
&\mp&\mbox{Tr}[{\tilde O}^{(i)}\Gamma_j^{(n)}1\sigma^y1{\tilde O}^{(i)}1\sigma^y1(\Gamma_j^{(n)})^T]-\mbox{Tr}[{\tilde O}^{(i)}\Gamma_j^{(n)}\tau^z\sigma^y1{\tilde O}^{(i)}\tau^z\sigma^y1(\Gamma_j^{(n)})^T]\}\frac{m^*}{4\pi}, \\
{\tilde B}^{(5/6)}_{ij}&=&-\tfrac{1}{32}\sum_{n=1}^{m_j}\{\mbox{Tr}[{\tilde O}^{(i)}\Gamma_j^{(n)}1\sigma^z1{\tilde O}^{(i)}1\sigma^z1(\Gamma_j^{(n)})^T]\pm\mbox{Tr}[{\tilde O}^{(i)}\Gamma_j^{(n)}\tau^z\sigma^z1{\tilde O}^{(i)}\tau^z\sigma^z1(\Gamma_j^{(n)})^T]\}. \label{Eq:Bt56}
\end{eqnarray}

The coefficients $b_i$ in Eq.\ \eqref{Eq:DAppFlow} are
\begin{equation}
b_i=b_i(\text{tadpole})+b_i(\text{sunrise}). \label{Eq:bCoeffs}
\end{equation}
Here, ``tadpole'' and ``sunrise'' refer to the corresponding diagrams considered in Ref.\
\onlinecite{CvetkovicPRB2012}.  Of the ``tadpole'' contributions, only $b_{A_1,2}(\text{tadpole})$
is non-zero, and is given by $8\times\frac{m^*}{4\pi}$.  The ``sunrise'' term contributes to all
of the coefficients, and is given by
\begin{equation}
b_i(\text{sunrise})=\tfrac{1}{8}\sum_{m}\mbox{Tr}(1\sigma_31\Gamma_i^{(m)} 1\sigma_31\Gamma_i^{(m)})\frac{m^*}{4\pi}.
\end{equation}

\end{widetext}

\section{Coefficients in the free energy} \label{App:FreeEnergyCoeffs}
We now list the coefficients $\alpha$ of the free energy, Eq.\ \eqref{Eq:FreeEnergy}.
The coefficients $\alpha_{a,i}^{\text{ph}}$ are
\begin{eqnarray}
\alpha_{1/2,i}^{\text{ph}}&=&8\pm\mbox{Tr}[(O^{(i)}\tau_{3}1_4)^2], \label{Eq:Alpha12PH} \\
\alpha_{3/4,i}^{\text{ph}}&=&-\tfrac{1}{2}\{\mbox{Tr}[(O^{(i)}1\sigma_11)^2]\mp\mbox{Tr}[(O^{(i)}\tau_3\sigma_11)^2] \cr
&\mp&\mbox{Tr}[(O^{(i)}1\sigma_21)^2]+\mbox{Tr}[(O^{(i)}\tau_3\sigma_21)^2]\} \\
\alpha_{5/6,i}^{\text{ph}}&=&\mbox{Tr}[(O^{(i)}1\sigma^z1)^2]\pm\mbox{Tr}[(O^{(i)}\tau^z\sigma^z1)^2].
\end{eqnarray}
The $\alpha_{a,i}^{\text{pp}}$ coefficients are
\begin{eqnarray}
\alpha_{1/2,i}^{\text{pp}}&=&8\mp\mbox{Tr}[({\tilde O}^{(i)}\tau_{3}1_4)^2], \\
\alpha_{3/4,i}^{\text{pp}}&=&\tfrac{1}{2}\{\mbox{Tr}[({\tilde O}^{(i)}1\sigma_11)^2]\pm\mbox{Tr}[({\tilde O}^{(i)}\tau_3\sigma_11)^2] \cr
&\mp&\mbox{Tr}[({\tilde O}^{(i)}1\sigma_21)^2]-\mbox{Tr}[({\tilde O}^{(i)}\tau_3\sigma_21)^2]\}. \\
\alpha_{5/6,i}^{\text{pp}}&=&\mbox{Tr}[(O^{(i)}1\sigma^z1)^2]\pm\mbox{Tr}[(O^{(i)}\tau^z\sigma^z1)^2]. \label{Eq:Alpha56PP}
\end{eqnarray}

\end{document}